\documentclass[aps,showpacs,pra,twocolumn,superscriptaddress,a4paper]{revtex4-1}
\usepackage{amsmath,amssymb}
\usepackage[utf8x]{inputenc}
\usepackage{textcomp,marvosym}
\usepackage{nameref,hyperref}

\usepackage{graphicx}

\usepackage{float}

\usepackage{cleveref}
\crefname{figure}{Fig.}{Figs} 
\Crefname{figure}{Fig.}{Figs} 
\crefname{section}{sec.}{sections} 
\Crefname{section}{Section}{Sections} 
\crefname{table}{Table}{Tables} 
\crefname{equation}{Eq.}{Eqs} 
\Crefname{equation}{Eq.}{Eqs} 
\crefname{appendix}{Appendix}{Appendices} 
\Crefname{appendix}{Appendix}{Appendices} 

\usepackage{microtype}
\DisableLigatures[f]{encoding = *, family = * }
\usepackage[table]{xcolor}

\usepackage{array}
\newcolumntype{L}{>{$}l<{$}} 

\newcolumntype{+}{!{\vrule width 2pt}}

\usepackage{epstopdf}
\newcommand{\be}{\begin{equation}}
\newcommand{\ee}{\end{equation}}

\begin{document}

\title{
Impact of climate change on backup energy and storage needs in wind-dominated power systems in Europe
}
\author{Juliane Weber}
\affiliation{Forschungszentrum J\"ulich, Institute of Energy and Climate Research -- Systems Analysis and Technology Evaluation (IEK-STE), 52425 J\"ulich, Germany}
\affiliation{University of Cologne, Institute for Theoretical Physics, Z\"ulpicher Str.~77, 50937 Cologne, Germany}	
	
\author{Jan Wohland}
\affiliation{Forschungszentrum J\"ulich, Institute of Energy and Climate Research -- Systems Analysis and Technology Evaluation (IEK-STE), 52425 J\"ulich, Germany}
\affiliation{University of Cologne, Institute for Theoretical Physics, Z\"ulpicher Str.~77, 50937 Cologne, Germany}	
	
\author{Mark Reyers}
\affiliation{University of Cologne, Institute for Geophysics and Meteorology, Pohligstr.~3, 50969 Cologne, Germany}
               
\author{Julia Moemken}
\affiliation{University of Cologne, Institute for Geophysics and Meteorology, Pohligstr.~3, 50969 Cologne, Germany}
\affiliation{Karlsruhe Institute of Technology, Institute of Meteorology and Climate Research, Wolfgang-Gaede-Strasse 1, 76131 Karlsruhe, Germany}

\author{Charlotte Hoppe}
\affiliation{Forschungszentrum J\"ulich, Institute of Energy and Climate Research -- Troposphere (IEK-8), 52425 J\"ulich, Germany}
\affiliation{University of Cologne, Rhenish Institute for Environmental Research, Aachener Str.~209, 50931 Cologne, Germany}	

\author{Joaquim G. Pinto}
\affiliation{Karlsruhe Institute of Technology, Institute of Meteorology and Climate Research, Wolfgang-Gaede-Strasse 1, 76131 Karlsruhe, Germany}

\author{Dirk Witthaut}
\affiliation{Forschungszentrum J\"ulich, Institute of Energy and Climate Research -- Systems Analysis and Technology Evaluation (IEK-STE), 52425 J\"ulich, Germany}
\affiliation{University of Cologne, Institute for Theoretical Physics, Z\"ulpicher Str.~77, 50937 Cologne, Germany}		

\date{\today }

\begin{abstract}
The high temporal variability of wind power generation represents a major challenge for the realization of a sustainable energy supply. Large backup and storage facilities are necessary to secure the supply in periods of low renewable generation, especially in countries with a high share of renewables. We show that strong climate change is likely to impede the system integration of intermittent wind energy. To this end, we analyze the temporal characteristics of wind power generation based on high-resolution climate projections for Europe and uncover a robust increase of backup energy and storage needs in most of Central, Northern and North-Western Europe. This effect can be traced back to an increase of the likelihood for long periods of low wind generation and an increase in the seasonal wind variability.
\end{abstract}

\maketitle

\section*{Introduction}

The mitigation of climate change requires a fundamental transformation of our energy system. Currently, the generation of electric power with fossil fuel-fired power plants is the largest source of carbon dioxide emissions with a share of approximately 35~\% of the global emissions \cite{Bruc14}. These power plants must be replaced by renewable sources such as wind turbines and solar photovoltaics (PV) within at most two decades to meet the 2$^\circ$C or even the 1.5$^\circ$C goal of the Paris agreement \cite{Paris15,roge15,Rogelj16}. While wind and solar power have shown an enormous progress in efficiency and costs \cite{Jacobson11,Sims11}, the large-scale integration into the electric power system remains a great challenge.

The operation of wind turbines is determined by weather and climate and thus strongly depends on the regional atmospheric conditions. Hence, the generated electric power is strongly fluctuating on different time scales. These fluctuations are crucial for system operation \cite{Bloo16,olau16,davi16,Sims11,Hube14,weber17}. In particular, large storage and backup facilities are needed to guarantee supply also during periods of low wind generation \cite{Heid10,Elsn15,diaz12}. How does climate change affect these fluctuations and the challenges of system integration? Previous studies have addressed the impact of climate change on the availability of cooling water \cite{Vlie12,Vlie16}, the energy demand \cite{Allen2016,Auffhammer2017}, or the change of global energy yields of wind and solar power \cite{pryo10,Tobi15,Tobi16,Reye15,Reye16,2017Moemken,Jere15,stan16}.
However, the potentially crucial impact of climate change on temporal wind fluctuations has not yet been considered in terms of system integration.

A consensus exists about general changes in the mean sea level pressure and circulation patterns in the European/North Atlantic region \cite{Demu09,Woll12,Zapp15,Lu07}. A projected increase of the winter storminess over Western Europe \cite{Pint12,Fese15} leads to enhanced wind speeds over Western and Central Europe, while in summer a general decrease is identified \cite{hueg13,Reye15,Reye16,2017Moemken}. This can lead to a strong increase of the seasonal variability of wind power generation and thus impede system integration, even though the annual mean changes are comparatively small. 

In this article, we study how climate change affects the temporal characteristics of wind power generation and the necessity for backup and storage infrastructures in wind-dominated power systems in individual European countries. Our analysis is based on five state-of-the-art global circulation models (GCMs) downscaled by the EURO-CORDEX initiative \cite{Jaco14,gior15}. We complement our results with an assessment of the large ensemble of the Coupled Model Intercomparison Project Phase 5 (CMIP5, \cite{Tayl12}) based on circulation weather types \cite{Jone93}. The paper is organized as follows. We first introduce our model to derive the backup need of a country as a function of the storage capacity. Additionally, we present the methods to analyze the CMIP5 ensemble. Afterwards we report our results. The article closes with a discussion.

\section*{Methods}

\begin{figure*}[tb]
\centering
\includegraphics[width=0.95\textwidth]{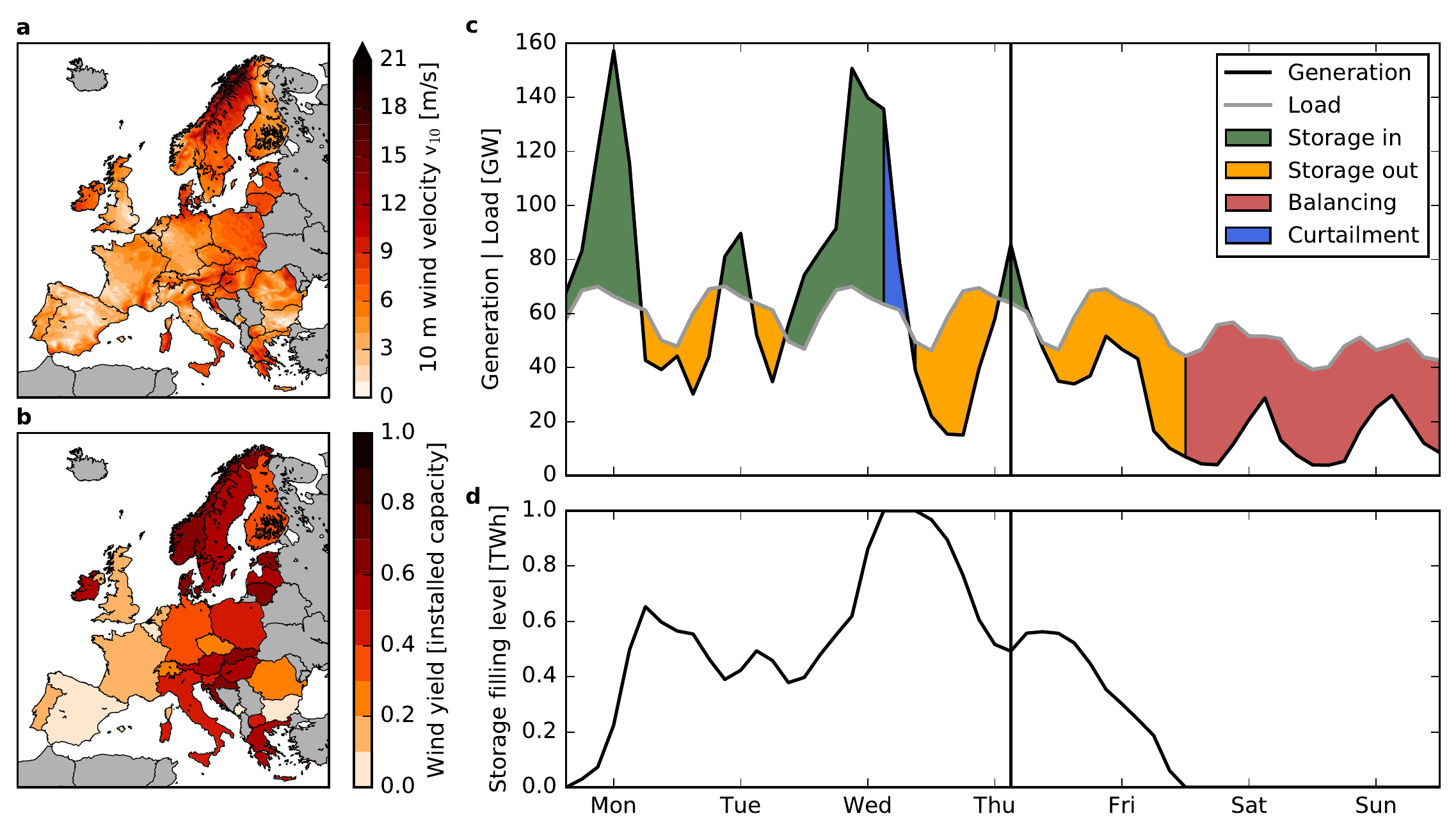}
\caption{
\label{Fig1}
{\bf Conversion of near-surface wind speeds to country-wise aggregated wind power generation combined with backup and storage infrastructures.} \textbf{a}, Near-surface wind velocities of the downscaled ERA-Interim data \cite{dee11} over Europe for one exemplary point in time. \textbf{b}, Corresponding estimated wind power yield for each country in units of the installed capacity. \textbf{c}, Renewable generation (black) and load (grey) in Germany for one exemplary week in spring assuming a power system with 100 \% wind power on average.
The vertical line denotes the time selected in panels \textbf{a} and \textbf{b}. The color indicates the operation of the storage system. Green: Excess power is stored. Yellow: Residual load is covered by the storage. Red: Residual load is covered by backup power plants as the storage is empty. Blue: Excess power must be curtailed as the storage is fully charged. \textbf{d}, Evolution of the storage filling level $S(t)$. 
}
\end{figure*}

The operation of future renewable power systems with large contributions of wind crucially depends on weather and climate. GCMs are used to simulate the dynamics of the earth system on coarse spatial scales for different scenarios of future greenhouse gas concentrations (representative concentration pathways, RCPs \cite{Vuur11}). To analyze the operation of the electric power system, a high spatial and temporal resolution is required. Our analysis is thus based on a subset of the EURO-CORDEX ensemble which provides dynamically downscaled climate change data at high resolution ($0.11^{\circ}$ and 3 hours). Time series for the aggregated wind power generation in a country are obtained from the near-surface wind speed (see \cref{Fig1}a, b).

Backup and storage infrastructures are needed when renewable generation drops below load. In order to quantify the necessary amount of backup and storage to ensure a stable supply, we adopt a coarse-grained model of the electric power system (see \cref{Fig1}c, d). Backup and storage needs crucially depend on the temporal characteristics of wind power generation, in particular the length of periods with low wind generation and the seasonal variability. In the present paper, we thus focus on temporal characteristics and their potential alteration due to climate change.

\subsection*{Wind power generation time series} \label{sec:Methods_power_timeseries}
Our analysis is based on a subset of the EURO-CORDEX regional climate simulations which provides dynamically downscaled climate change data at high resolution for Europe based on five GCMs: CNRM-CM5, EC-EARTH, HadGEM2-ES, IPSL-CM5A-MR, MPI-ESM-LR \cite{Jaco14} (see also \cref{tbl:GCMs} in \cref{sec:a1}). All data is freely available for example at the ESGF (Earth System Grid Federation) node at DKRZ (German Climate Computing Centre) \cite{esgf}. The five models are downscaled using the hydrostatic Rossby Centre regional climate model RCA4 \cite{Stra15,Samu11}. The downscaling provides continuous surface (10~m) wind data from 1970 to 2100 with a spatial resolution of $0.11^{\circ}$ and a temporal resolution of $T = $ 3 hours. Unfortunately, downscaled data at this high spatial and temporal resolution is not yet available for more GCMs or for different regional climate models at ESGF \cite{esgf}. Considering the use of only one regional climate model, \cite{2017Moemken} show that differences between different GCMs are usually larger than differences between different regional climate models. 

We analyze a strong climate change scenario (RCP8.5) using a rising radiative forcing pathway leading to additional 8.5~W/m$^2$ ($\sim$1370~ppm CO$_2$ equivalent) by 2100 and a medium climate change scenario (RCP4.5, $\sim$650~ppm CO$_2$ equivalent, see \cref{sec:a3}) \cite{Vuur11}. We compare two future time frames, 2030-2060 (mid century, `mc') and 2070-2100 (end of century, `eoc'), to a historical reference time frame (1970-2000, `h').

The calculation of wind power generation requires wind speeds at the hub height of wind turbines. As the high resolution wind velocities are only available at a height of $z_0 = 10$~m, they must be extrapolated to a higher altitude. We choose a hub height of $z=90$~m as in \cite{Tobi16} and extrapolate the surface wind velocities $v_{z_0}$ using a power law formula: $v_z = v_{z_0} \left( {z}/{z_0}\right)^{1/7}$ \cite{Manwell09}. Although widely used, this simple formula is only valid for smooth open terrain and only applies for a neutrally stable atmosphere. Unfortunately, the available data set does not allow to assess the stability of the atmosphere. Thus, it is unclear how to improve the scaling law with the present data available. Tobin \textit{et al.} \cite{Tobi16} show in a sensitivity study that their results hardly depend on the extrapolation technique or on the chosen hub height. They further state that the ``uncertainty related to climate model formulation prevails largely over uncertainties lying in the methodology used to convert surface wind speed into power output''.

The wind generation is derived using a standardized power curve with a cut-in wind speed of $v_i = 3.5$ m/s, a rated wind speed of $v_r = 12$ m/s and a cut-out wind speed of $v_o = 25$ m/s as in \cite{Tobi16}. The capacity factor $CF(t)$ (i.e. the generation normalized to the rated capacity) then reads
\be
CF(t) =
\left\{ \begin{array}{@{\kern2.5pt}lL}
    \hfill 0 & if $v_z(t) < v_i$ or $v_z(t) \geq v_o$.\\
    \hfill \frac{v_z^3(t)-v_i^3}{v_r^3-v_i^3} & if $ v_i \leq v_z(t)$,\\
    \hfill 1 & else.
\end{array}\right.
\ee
In order to account for wind farms and velocity variations, the power curve is smoothed using a gaussian kernel (see \cref{fig:Power_Curve} in \cref{sec:a1} and \cite{Andresen15, Manwell09}).

To obtain the gross generation per country, we equally distribute wind farms on grid points for which the local average wind yield is higher than the country average (see \cref{fig:CF_weights} in \cref{sec:a1}) \cite{Monf16}. Offshore sites are not considered as we want to focus on effects of climate change on onshore wind power generation in this study. The distribution is fixed using historical reanalysis data from ERA-Interim \cite{dee11} downscaled by the EURO-CORDEX initiative \cite{Jaco14,Stra15} to guarantee consistency (cf.~\cref{Fig1}b). We do not use the wind farm distribution as of today because installed capacities in a fully-renewable power system will be much higher and also more widespread than they are today such that wind parks will be built in yet unused locations. Furthermore, \cite{Monf16} show that different wind farm distributions do not significantly affect the results (see also \cref{fig:increase-backup_no_weights,fig:longcalms_no_weights,fig:seasonality_no_weights} in \cref{sec:a4} where we tested a homogeneous wind farm distribution within each country).

Wind power generation is aggregated using two approaches: (a) aggregation per country neglecting transmission constraints, assuming an unlimited grid within each country; (b) aggregation over the whole European continent, assuming a perfectly interconnected European power system (copperplate). The intermediate case is discussed for instance in \cite{Rodriguez2014,Schlachtb16} for current climatic conditions and in \cite{2017Wohland} for a changing climate but without considering storage.

For the load time series $L(t)$ we use data of the year 2015 provided by the European Network of Transmission System Operators for Electricity (ENTSO-E, \cite{entsoe_load}) and repeat this year 31 times. In order to avoid trends in the load timeseries, we consider a single year only. Furthermore, we show in a sensitivity study assuming constant loads that our results dominantly depend on the generation timeseries and are hardly affected by the load time series (see \cref{fig:increase-backup_real_demand} in \cref{sec:a4}). Throughout all time frames we assume that wind power provides a fixed share $\gamma$ of the load $L(t)$ per country \cite{Heid10}. Hence, the fluctuating wind power generation $R(t)$ is scaled such that
\be
R(t) = \gamma \, \frac{CF(t)}{\langle CF(t) \rangle} \, \langle L(t) \rangle, 
\ee

\noindent where the brackets denote the average over the respective time frame for a given model. This procedure normalizes out a possible change of the gross wind power yield, and thus allows to isolate the effects of a change in the temporal distribution of the renewable generation. For the copperplate assumption, the wind power generation is scaled such that each country provides a fixed share $\gamma$ of the country-specific load. Afterwards, the country-specific wind power generation is summed-up to one aggregated time series. In the main manuscript, we focus on a fully renewable power system per country, i.e.,~$\gamma = 1$, with 100 \% wind power generation. Results for different values of $\gamma$ are shown in \cref{fig:increase-backup_gamma120,fig:increase-backup_gamma80} in \cref{sec:a4}.

\subsection*{Calculation of backup energy needs} \label{sec:Methods_balancing}
Country-wise aggregated wind generation and load data are used to derive the backup energy need of a country given different storage capacities. At each point of time $t$ power generation and consumption of a country must be balanced  \cite{Heid10,Rasm12,Rodriguez2014}

\be
   R(t) + B(t) = \Delta(t) + L(t) + C(t),
   \label{eq:powerbalance}
\ee
where $R(t)$ and $B(t)$ denote the generation by fluctuating renewables and dispatchable backup generators, respectively, $L(t)$ is the load and $C(t)$ denotes curtailment (cf.~\cref{Fig1}c). $\Delta(t)$ is the generation ($\Delta(t) < 0$) or load ($\Delta(t) > 0$) of the storage facilities, such that the storage filling level evolves according to (cf.~\cref{Fig1}d)
\be
S(t+T) = S(t) + \Delta(t) \cdot T.
\ee
where $T$ is the duration of one time step (here: 3 hours). The storage filling level must satisfy $0 \le S(t) \le S_{\rm max}$ with $S_{\rm max}$ being the storage capacity. We decide to minimize the total backup energy $B_{\rm tot} = \sum_t B(t) \cdot T$ which also minimizes fossil-fuel usage and hence greenhouse gas emissions. This yields a storage-first strategy \cite{Rasm12}: In the case of overproduction (i.e. $R(t) > L(t)$) excess energy is stored until the storage device is fully charged,
\be
   \Delta(t) = \min [ R(t)-L(t) ; (S_{\rm max} - S(t))/T ].
\ee
To ensure power balance, we may need curtailment 
\be
C(t)=R(t)-L(t)-\Delta(t).
\ee
In the case of scarcity (i.e. $R(t) < L(t)$) energy is provided by the storage infrastructures until they are empty,
\be
   \Delta(t) = - \min [ L(t)-R(t) ; S(t)/T ].
\ee
The missing energy has to be provided by backup power plants, 
\be
B(t)=L(t)-R(t)+\Delta(t).
\ee
The backup power $B$ is not restricted in our model and can be interpreted as the aggregated amount of backup power per country, not differentiating between different technologies. In order to keep the storage neutral, a periodic boundary condition ($S(t=t_{\rm max}) = S(t=0)$) is applied \cite{Jens14}. We emphasize that by the term `storage' we mean storage regardless of the technical realization. Hence, $S_{\rm max}$ describes the total accumulated storage capacity, including virtual storage. For simplicity, we neglect losses in the storage process.

In the figures we show the average backup energy per year $E = \langle B \rangle / \langle L \rangle \cdot L_{\rm year}$. $L_{\rm year}$ is the average yearly gross electricity demand of the respective country. Thus, $E/L_{\rm year}$ gives the share of energy that has to be provided by dispatchable backup generators \cite{Rasm12}.

\subsection*{Persistence of low wind situations} \label{sec:Methods_duration}
We measure the probability for long low wind periods during which a high amount of energy is required from storage devices and backup power plants. Therefore, we identify all periods for which the wind power generation is continuously smaller than average (i.e. $R(t) < \langle R \rangle$) and record their duration $\tau$. From this, we can estimate a probability distribution. Extreme events are quantified by the 95 \% quantile of the distribution.

\subsection*{Seasonal wind variability} 
The wind yield in Europe is usually higher in winter than in summer. An increasing seasonal wind variability would refer to higher wind yields in the winter months and/or lower wind yields in the summer months and would lead to higher backup energy needs during summer. 

We define the winter-summer ratio of the country-wise aggregated wind power generation as the ratio of the average winter wind generation $\langle R \rangle_{\rm DJF}$ and the average summer wind generation $\langle R \rangle_{\rm JJA}$:
\be
	R_{\rm winter-summer} = \frac{\langle R \rangle_{\rm DJF}}{\langle R \rangle_{\rm JJA}},
\ee

\noindent with `DJF': December, January, February, and `JJA': June, July, August. $\langle R \rangle_{\rm DJF}$ and $\langle R \rangle_{\rm JJA}$ are the mean generations within a certain time frame (historical, mid century, end of century).

\subsection*{Analysis of low wind periods using a statistical analysis of a large CMIP5 ensemble} \label{sec:Methods_CWTs}

Our analysis is complemented with lower resolution data of 22 GCMs contributing to the Coupled Model Intercomparison Project Phase 5 (CMIP5, \cite{Tayl12}). The GCM output is analyzed with a statistical method developed in \cite{Reye15,Reye16}. We characterize the large-scale circulation over Central Europe by determining the prevalent circulation weather type (CWT, \cite{Jone93}) using instantaneous daily mean sea level pressure (MSLP) fields around a central point at 10$^\circ$E and 50$^\circ$N (near Frankfurt, Germany) at 00 UTC (see also Fig~2 in \cite{Reye15}). The different CWT classes are either directional (`North', `North-East', `East', `South-East', `South', `South-West', `West', `North-West') or rotating (`Cyclonic', `Anti-cyclonic'). Additionally, a proxy for the large-scale geostrophic wind (denoted as $f$-parameter) is derived using the gradient of the instantaneous MSLP field. Higher geostrophic wind values (i.e. higher $f$-parameters) correspond to larger wind power yields in Central Europe.

In order to compare the CMIP5 and the EURO-CORDEX data, we test whether the $f$-parameter derived using the coarse ERA-Interim reanalysis data \cite{Reye15} is capable to reproduce the characteristics of German low-wind generation periods as determined from the downscaled ERA-Interim dataset \cite{dee11}. We classify days with below-average wind power generation (scarcity) for each CWT by a low value of the $f$-parameter, $f(t) \le f_{\rm th}$, as shown in \cref{Fig7}a. Thus, for each day, we can analyze whether the classifier ($f(t) \le f_{\rm th}$) correctly predicts that the wind power generation is below average ($R(t) < \langle R \rangle$) or erroneously predicts that the wind power generation is above average ($R(t) \ge \langle R \rangle$). The quality of this classification is quantified by the fraction of true predictions, called sensitivity
\be
   {\rm SEN} = \frac{n[{R < \langle R \rangle    \, \& \,  f \le f_{\rm th}}]}{
                             n[{R < \langle R \rangle    \, \& \, f \le f_{\rm th}}] 
                          + n[{R < \langle R \rangle \, \& \,  f > f_{\rm th}}]} 
\ee
and the fraction of false predictions
\be
   {\rm FFP} = \frac{ n[{R \ge \langle R \rangle \, \& \,  f \le f_{\rm th}}]}{
                             n[{R \ge \langle R \rangle \, \& \, f \le f_{\rm th}}] 
                          + n[{R \ge \langle R \rangle \, \& \,  f > f_{\rm th}}]}, 
\ee
where $n$ denotes the number of days where the conditions are satisfied \cite{Egan1975}.
A compromise must be found between a maximum sensitivity for high values of $f_{\rm th}$ and a minimum fraction of false predictions for low values of $f_{\rm th}$. A common choice is to choose the value $f_{\rm th}$ which minimizes $(1-{\rm SEN})^2 + {\rm FFP}^2$ \cite{Egan1975} (see also \cref{fig:ROC} in \cref{sec:a1} (ROC-curve)). Under the assumption that the meaning of the $f$-parameter does not depend on GCM and time frame, we use the derived $f_{\rm th}$ to estimate the duration of low wind periods as described in above.

\section*{Results}\label{sec:Results}

\begin{figure*}[tb]
\centering
\includegraphics[width=0.95\textwidth]{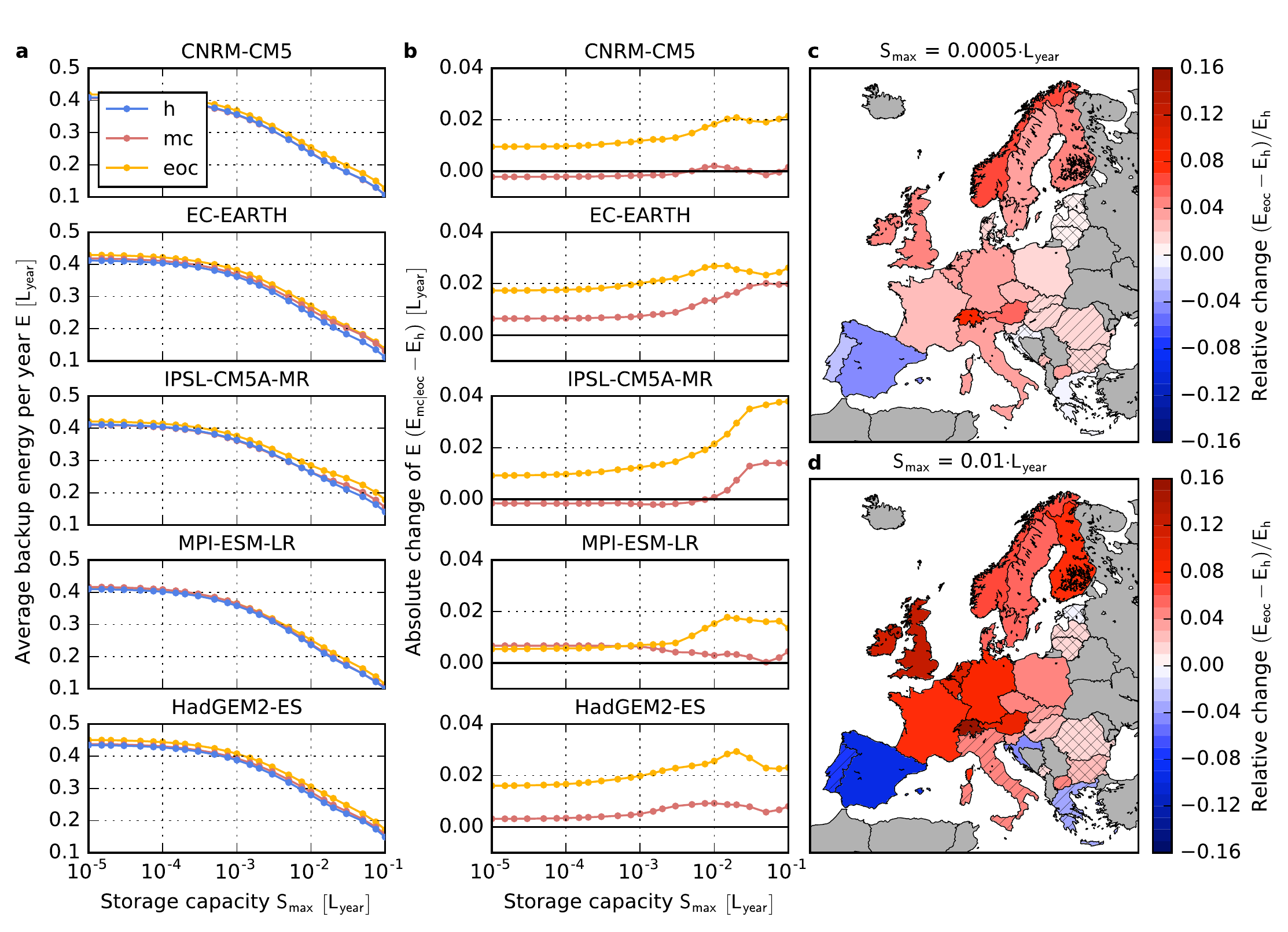}
\caption{
\label{Fig2}
{\bf Impact of strong climate change on backup energy needs in Europe.} \textbf{a,} Amount of energy that has to be provided by dispatchable backup generators in Germany as a function of the storage capacity $S_{\rm max}$ for the five models in the EURO-CORDEX ensemble and a strong climate change scenario (RCP8.5). Energy is given in units of the average yearly gross electricity consumption $L_{\rm year}$. Blue: 1970-2000 (h), Red: 2030-2060 (mc), Yellow: 2070-2100 (eoc). 
\textbf{b,} Absolute change of the average backup energy as a function of $S_{\rm max}$ in Germany. 
Colors are the same as in panel \textbf{a}. 
\textbf{c, d,} Relative change of the average backup energy needs by the end of the century with respect to the historical time frame for 29 European countries and two values of the storage capacity $S_{\rm max}$. The color code corresponds to the average of the five models and the hatching indicates the robustness of the results. No hatching: 5/5, striped: 4/5, crossbred: 3/5 models agree on the sign of change.
}
\end{figure*}

\subsection*{Increase of backup and storage needs}

All models in the EURO-CORDEX ensemble predict an increase of the necessary backup energy in most of Central Europe (i.e.~Germany, Poland, Czech Republic, Switzerland, Austria, the Netherlands and Belgium), France, the British Isles and Scandinavia for a strong climate change scenario (RCP8.5) by the end of the century relative to the historical time frame (see \cref{Fig2}c and d). This implies that even though the same amount of energy is produced by renewables in both time frames, less renewable energy can actually be used. Relative changes are highest in Switzerland and the United Kingdom with a range between 12.2 to 24.2 \% (mean: 15.6 \%) and 7.1 to 16.5 \% (mean: 12.1 \%), respectively for a storage capacity of $S_{\rm max} = 0.01 \cdot L_{\rm year}$. However, results for mountainous regions like Switzerland should be regarded with caution as wind farms might be placed at sites which are unsuitable. In addition, climate model results over complex terrain are known to have large uncertainties. An opposite effect is observed for the Iberian Peninsula, Greece and Croatia where the need for backup energy decreases (e.g.,~Spain: -4.7 to -15.5 \%; mean: -9.1 \% for $S_{\rm max} = 0.01 \cdot L_{\rm year}$). These results hold for a variety of scenarios for the development of storage infrastructures leading to different values of the storage size $S_{\rm max}$, being more pronounced for larger storage sizes. The latter partly results from a change in the seasonal variability of the wind power generation (cf.~below). In the Baltic region and South-Eastern Europe, relative changes are weaker and the models most often do not agree on the sign of change.

Similar changes are observed already at mid century (2030-2060, see \cref{fig:increase-backup_mc} in \cref{sec:a2}) and for RCP4.5 (see \cref{fig:increase-backup_rcp45_eoc} in \cref{sec:a3}). However, the results are less pronounced and often not robust (i.e. the models do not agree on the sign of change).

For Germany (\cref{Fig2}a and b), the absolute increase of the average backup energy per year $E$ amounts to 0.6-3.8~\% of the average yearly consumption $L_{\rm year}$ by the end of the century. Assuming a yearly consumption of the order of $L_{\rm year} = 600$ TWh \cite{AGE2016}, this corresponds to an additional need of 4-23 TWh of backup energy per year.

\begin{figure}[tb]
\centering
\includegraphics[width=0.95\columnwidth]{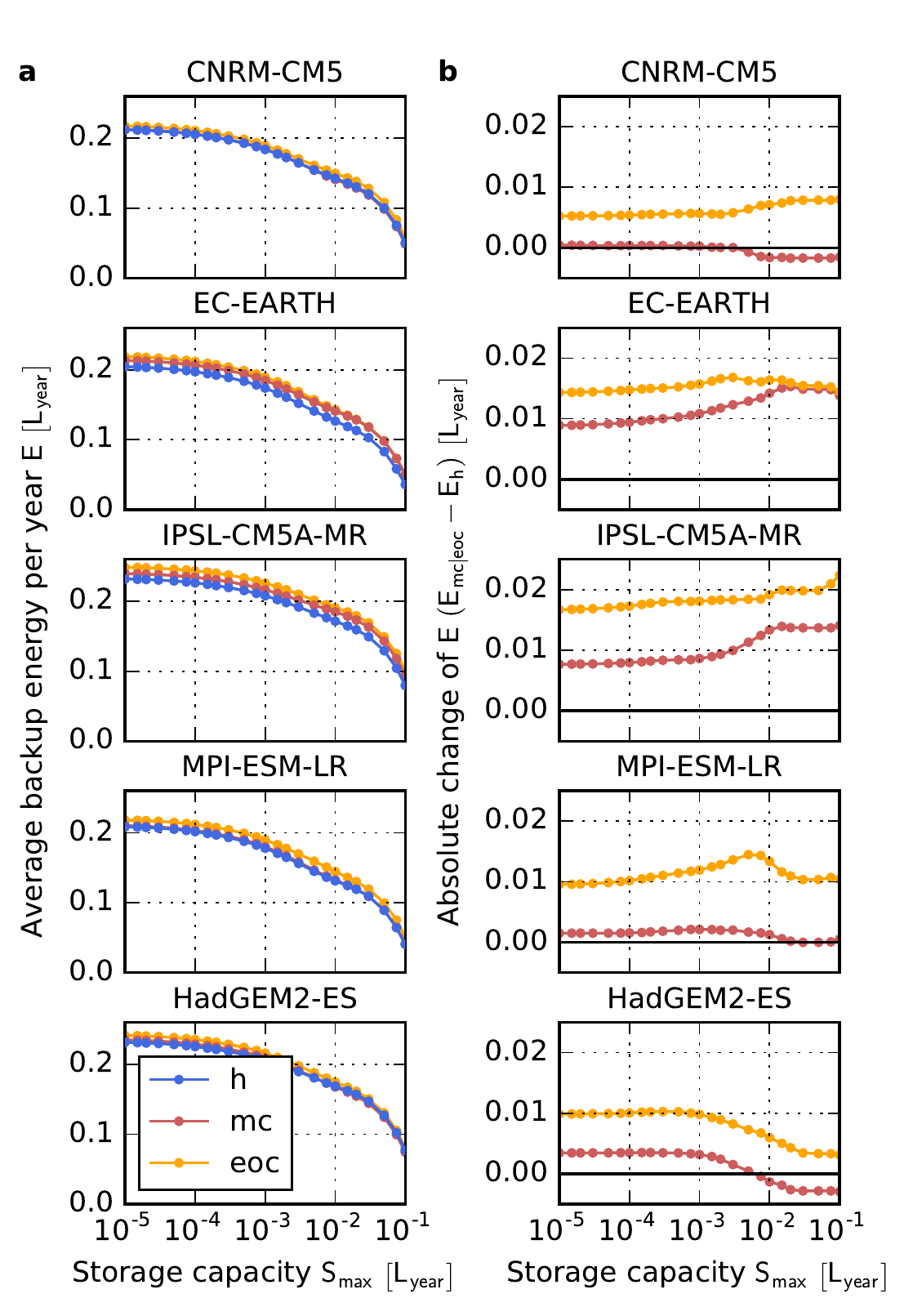}
\caption{
\label{Fig3}
{\bf Impact of strong climate change on backup energy needs for a perfectly interconnected European power system.} \textbf{a,} Amount of energy that has to be provided by dispatchable backup generators in Europe as a function of the storage capacity $S_{\rm max}$ for the five models in the EURO-CORDEX ensemble and a strong climate change scenario (RCP8.5). Energy is given in units of the average yearly gross electricity consumption $L_{\rm year}$. Blue: 1970-2000 (h), Red: 2030-2060 (mc), Yellow: 2070-2100 (eoc). 
\textbf{b,} Absolute change of the average backup energy as a function of $S_{\rm max}$. 
Colors are the same as in panel \textbf{a}. 
}
\end{figure}

In a perfectly interconnected Europe, the average relative  backup energy per year is much smaller than for individual countries (e.g.,~for Germany, cf~\cref{Fig2}a and \cref{Fig3}a). This is because the balancing takes place over a large spatial scale with many different wind patterns at the same time step. For all five models and all storage capacities, we find an increase of the average backup energy per year $E$ by the end of the century (see \cref{Fig3}). Values range from 0.3 to 2.2 \% of the average yearly consumption $L_{\rm year}$ corresponding to a relative change of 2.3 to 40.2 \% ($(E_{\rm eoc}-E_{\rm h})/E_{\rm h}$). For high storage capacities, the change depends strongly on the seasonal wind variability (cf.~below). For mid century, the same effect albeit at a weaker magnitude can be observed.

Two main drivers for the increase in the backup energy can be identified: a higher probability for long periods with low wind power generation and a higher seasonal wind variability.

\subsection*{Challenges by long low-wind periods}

\begin{figure*}[tb]
\centering
\includegraphics[width=0.95\textwidth]{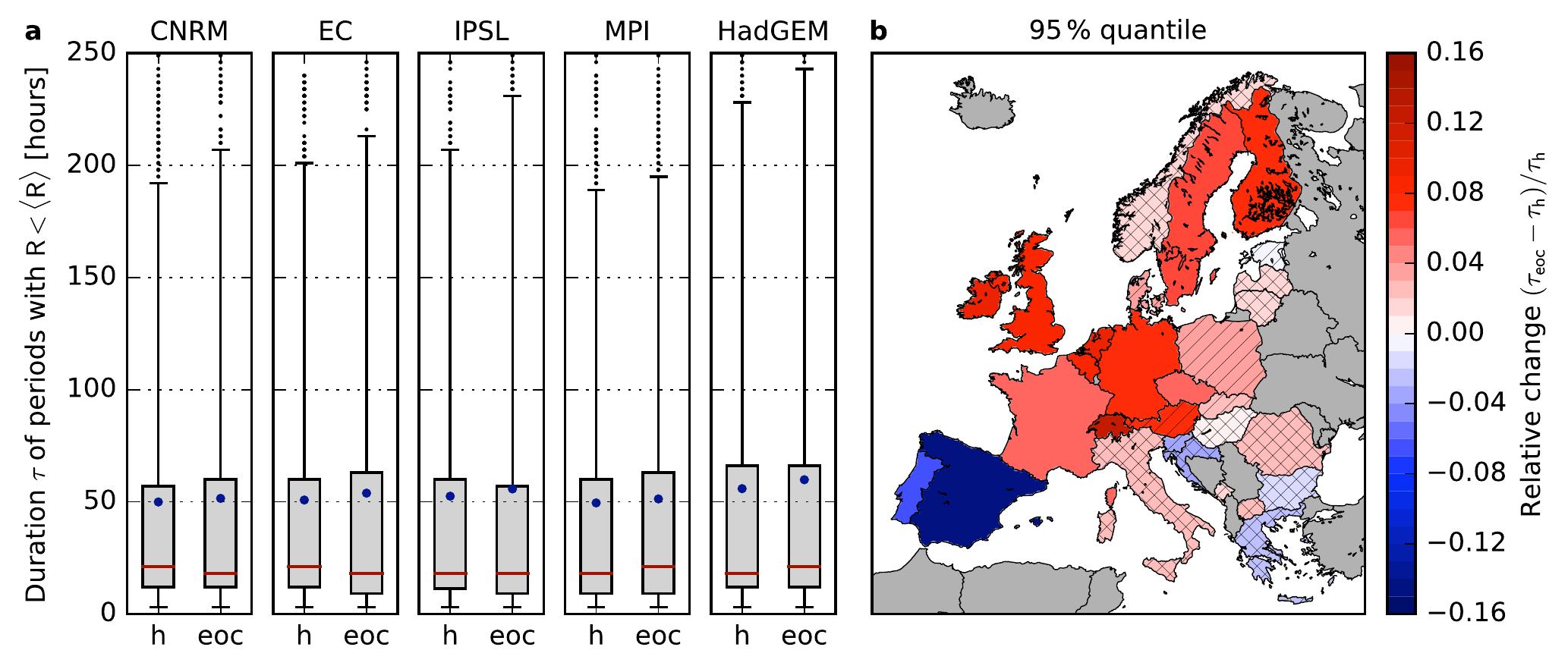}
\caption{
{\bf Change of the duration of periods with low wind generation.} 
\textbf{a}, Distribution of the duration of periods during which the wind generation is continuously lower than average ($R(t) < \langle R \rangle$) in Germany for the five models in the EURO-CORDEX ensemble. Boxes represent the 25~\% to 75~\% quantiles, whiskers indicate the 5~\% and 95~\% quantiles, the red line is the median, the blue dot shows the mean and black dots represent outliers. Results are shown for the historical time frame (h, 1970-2000) and the end of the century (eoc, 2070-2100) for a strong climate change scenario (RCP8.5). \textbf{b}, Relative change of the duration assigned to the 95~\% quantile by the end of the century with respect to the historical time frame for 29 European countries. The color code corresponds to the average of the five models and the hatching indicates the robustness of the results. No hatching: 5/5, striped: 4/5, crossbred: 3/5 models agree on the sign of change.
\label{Fig4}
}
\end{figure*}

The backup energy need is particularly high during long periods with low renewable generation which cannot be covered by limited storage facilities. In \cref{Fig4} we show the duration distribution of periods for which wind power generation is continuously lower than average (i.e. $R(t) < \langle R \rangle$) for Germany (panel a) and the relative change of the 95~\% quantile (panel b). The 95~\% quantile shifts to longer durations in most of Central Europe, France, the British Isles, Sweden and Finland and decreases on the Iberian Peninsula by the end of the century. These findings are robust in the sense that all five models in the EURO-CORDEX ensemble agree on the sign of change as illustrated for Germany in \cref{Fig4}a. 

Long low-wind periods are crucially difficult for the operation of future renewable power systems \cite{Elsn15}. An increasing magnitude for such extreme events thus represents a serious challenge for renewable integration. In Eastern Europe, Italy, Greece and Norway relative changes are weaker and not robust. The effect develops mostly in the second half of the century (cf.~\cref{fig:longcalms_mc} in \cref{sec:a2}) and for strong climate change (RCP8.5, cf.~\cref{fig:longcalms_rcp45_eoc} in \cref{sec:a3}).

\begin{figure}[tb]
\centering
\includegraphics[width=0.95\columnwidth]{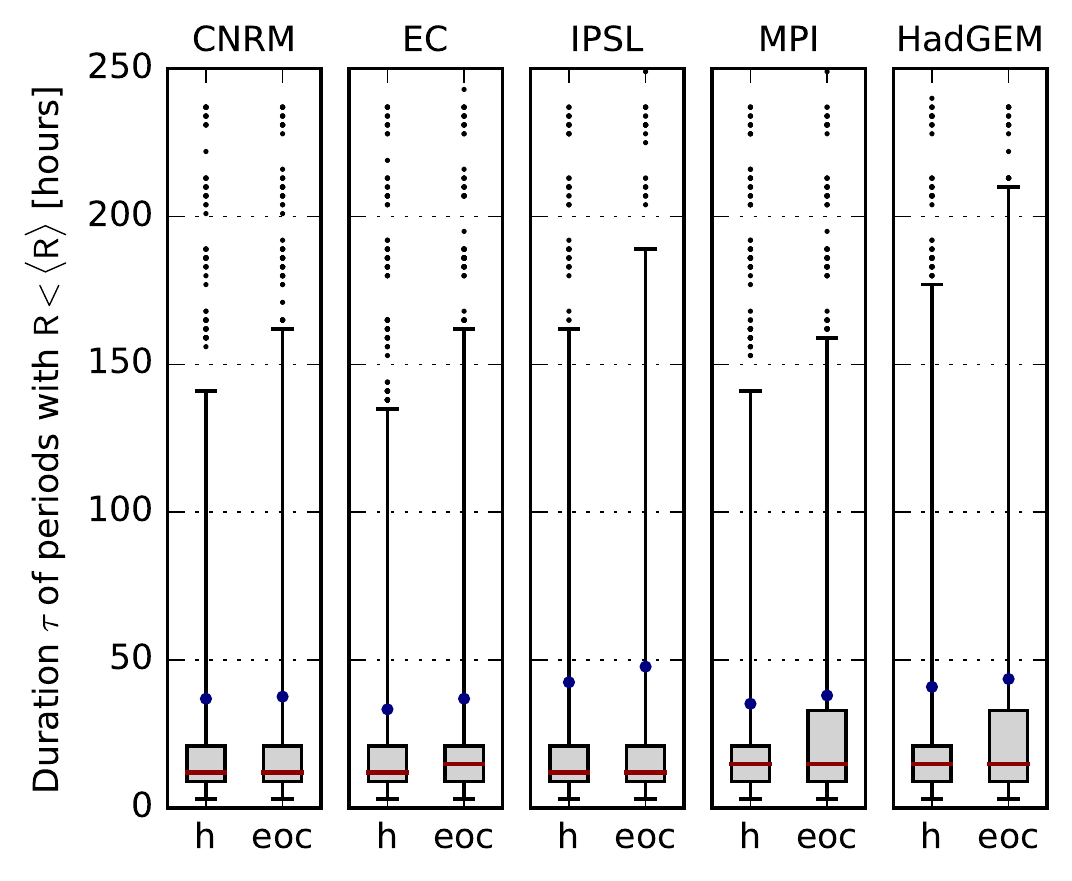}
\caption{
{\bf Change of the duration of periods with low wind generation for Europe as a whole.} Distribution of the duration of periods during which the wind generation is continuously lower than average ($R(t) < \langle R \rangle$) in Europe for the five models in the EURO-CORDEX ensemble. Boxes represent the 25~\% to 75~\% quantiles, whiskers indicate the 5~\% and 95~\% quantiles, the red line is the median, the blue dot shows the mean and black dots represent outliers. Results are shown for the historical time frame (h, 1970-2000) and the end of the century (eoc, 2070-2100) for a strong climate change scenario (RCP8.5).
\label{Fig5}
}
\end{figure}

Considering the perfectly interconnected European power system (see \cref{Fig5}), we find that the average duration of low-wind periods ($R(t) < \langle R \rangle$) and the 95 \% quantile both shift to higher values by the end of the century. This indicates that long lasting low-wind conditions which extend over the whole European continent become more likely (also discussed in \cite{2017Wohland}).

\subsection*{Higher seasonal wind variability} \label{sec:Results_variability}

\begin{figure}[tb]
\centering
\includegraphics[width=0.95\columnwidth]{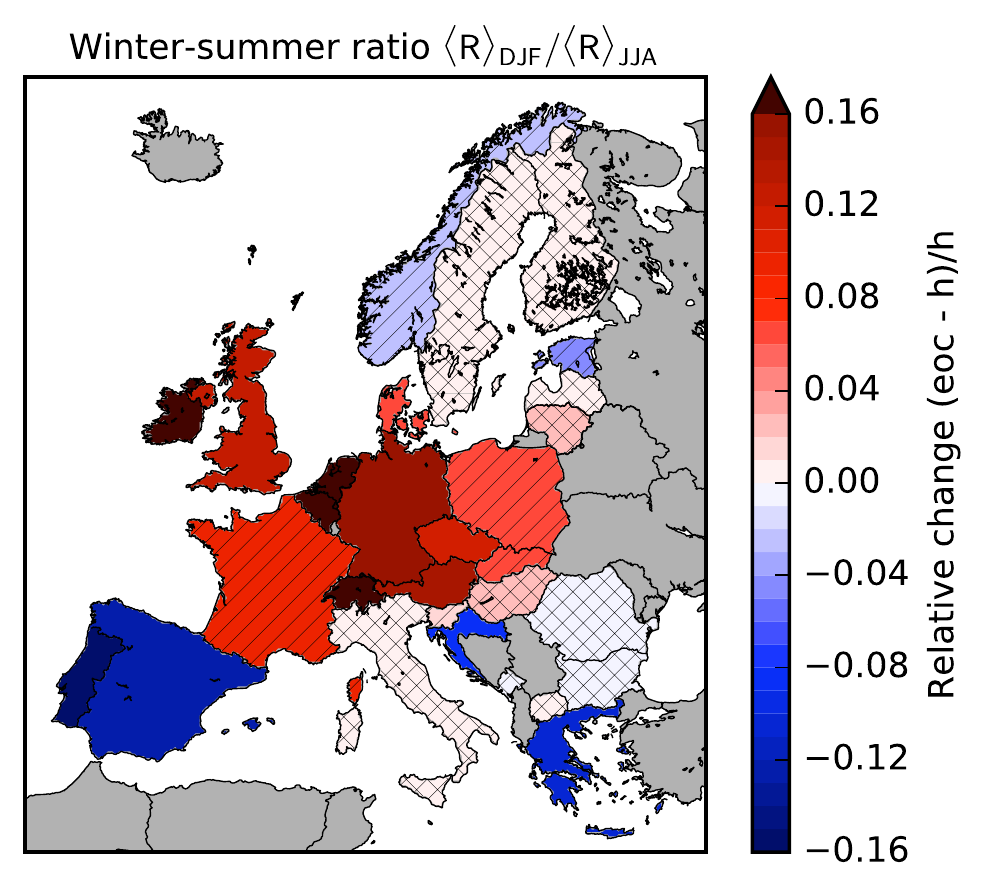}
\caption{
\label{Fig6}
{\bf Impact of strong climate change on the seasonal variability of wind power generation.} Relative change of the winter-summer ratio of the average wind power yield $\langle R \rangle_{\rm DJF}/\langle R \rangle_{\rm JJA}$ (DJF: December-February vs.~JJA: June-August) by the end of the century (eoc, 2070-2100) with respect to the historical time frame (h, 1970-2000). The brackets denote the temporal average over the respective winter and summer months. Results are shown for a strong climate change scenario (RCP8.5) for 29 European countries. The color code corresponds to the average of the five models and the hatching indicates the robustness of the results. No hatching: 5/5, striped: 4/5, crossbred: 3/5 models agree on the sign of change.
}
\end{figure}

The second reason for an increase of backup and storage needs is an increasing intensity of the seasonal wind variability. Typically, the wind power yield is highest in the winter months such that backup power plants are needed mostly in summer. 

The winter-summer ratio increases for most of Central and North-Western Europe, and decreases for the Iberian Peninsula, Greece and Croatia (see \cref{Fig6}) for four or all five models in the EURO-CORDEX ensemble. In these countries the seasonal variability therefore contributes to the observed changes of backup needs. Changes are small and not robust in Italy, most of Eastern Europe and Scandinavia (except Denmark). Hence, the increase of backup needs in Northern Europe is attributed solely to the higher probability for long periods with low wind power generation. For mid century (see \cref{fig:seasonality_mc} in \cref{sec:a2}), and for medium climate change (RCP4.5, see \cref{fig:seasonality_rcp45_eoc} in \cref{sec:a3}), results are comparable but less robust for some countries.

For the perfectly interconnected European power system, four of the five models predict an increasing seasonal wind variability in the range of 4.1 to 10.4 \%. Thus, the lower seasonal wind variability on the Iberian Peninsula, Greece and Croatia cannot totally compensate the higher seasonal wind variability in the other European countries. In contrast, HadGEM2-ES predicts a decrease of -2.8 \%.

The higher seasonal wind variability also explains the relative increase of the backup energy for higher storage capacities (cf.~\cref{Fig2}). A high storage capacity allows to store some part of the energy for several months. However, as the storage capacity is still limited, a higher seasonal wind variability implies that the storage is fully charged earlier in winter and that it is depleted earlier in summer. Thus, less excess energy can be transferred from the winter to the summer months if the seasonal variability of wind power generation increases.

\subsection*{Low-wind periods in a large CMIP5 ensemble}

To substantiate our findings, we analyze a large CMIP5 ensemble \cite{Tayl12} consisting of 22 GCMs with a much coarser resolution than the EURO-CORDEX ensemble as explained in the methods section. 

\begin{figure*}[tb]
\centering
\includegraphics[width=0.95\textwidth]{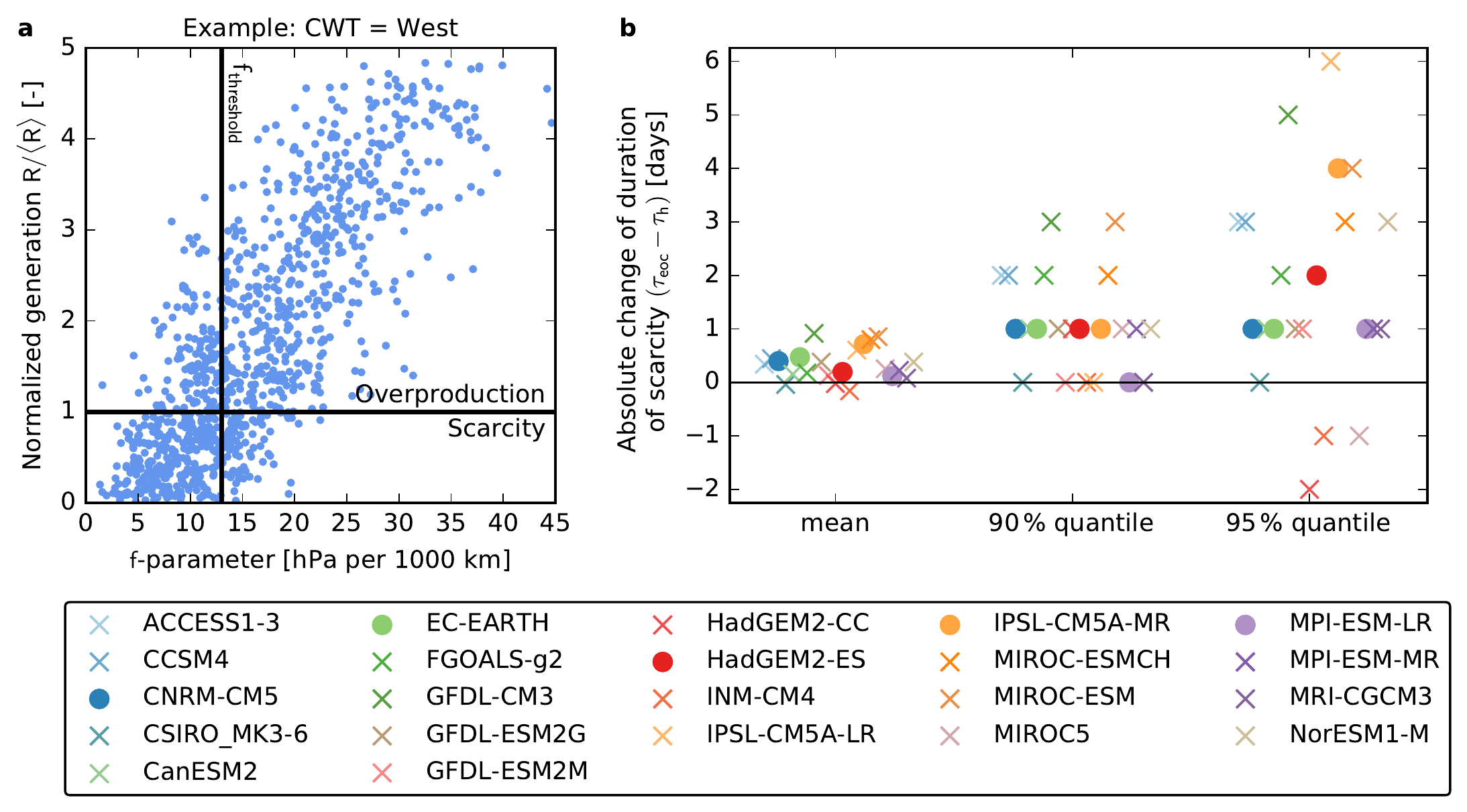}
\caption{
\label{Fig7}
{\bf Assessment of long low-wind periods in a large CMIP5 ensemble.} \textbf{a,} Days with below average wind power generation in Central Europe are identified by a low value of the $f$-parameter ($f(t) \le f_{\rm th}$) in the GCM output. To determine the optimal value of the threshold $f_{\rm th}$, for each circulation weather type (CWT, here, the western type is shown) we compare the $f$-parameter to the German wind power output $R(t)$, calculated from the dynamically downscaled ERA-Interim reanalysis dataset \cite{dee11}.
\textbf{b,} Absolute change of the duration of periods with $f(t) \le f_{\rm th}$ by the end of the century (eoc, 2070-2100) compared to the historical time frame (h, 1970-2000) for a strong climate change scenario (RCP8.5) for 22 GCMs in the CMIP5 ensemble. The change of the mean duration, the 90~\% quantile and the 95 \% quantile of the duration distribution are shown. Filled circles represent the five GCMs which are also downscaled by the EURO-CORDEX initiative.
}
\end{figure*}

The typical duration of periods with $f(t) \le f_{\rm th}$ in Central Europe increases by the end of the century for most GCMs in the CMIP5 ensemble. 19 of the 22 models predict an increase of the mean duration (\cref{Fig7}b). The 90~\% quantile of the duration increases for 16 models and remains unchanged for the remaining six models, while the 95~\% quantile increases for 18 of the 22 models. \Cref{Fig7}b shows that the five models of the EURO-CORDEX ensemble (shown as filled circles) form a representative subset of the CMIP5 ensemble since their results are well distributed within the range of the majority of all models and thus do not contain outliers. Hence, the large CMIP5 ensemble corroborates our previous findings, predicting an increase of the likelihood for long periods with low wind power output for a strong climate change scenario. 

To assess the sensitivity of the choice of $f_{\rm th}$, we repeated our analysis by determining one value for $f_{\rm th}$ which is independent of the underlying CWT. This does not change the results as shown in \cref{fig:CWT-findings_one_f_thr} in \cref{sec:a4}.

\subsection*{Climatologic developments driving enhanced seasonality}

The identified increase in the seasonal variability of wind power generation has been discussed in terms of the projected changes of large-scale atmospheric circulation and regional wind conditions. A consensus exists about general changes in the large-scale circulation patterns in the Eastern North Atlantic region and Europe, which is however dependent on the time of the year \cite{Demu09}. During winter, the eddy-driven jet stream and cyclone intensity are extended towards the British Isles \cite{Woll12}. Accordingly, winter storminess is projected to increase over Western Europe \cite{Pint12,Fese15}, leading to enhanced winds over Western and Central Europe. The signal in summer corresponds rather to a northward shift of the eddy driven jet stream, cyclone activity and lower tropospheric winds, together with an increase in anticyclonic circulation over Southern Europe \cite{Zapp15}. The latter is associated with an expansion of the Hadley circulation due to enhanced radiative forcing \cite{Lu07}. These developments are projected to decrease wind speeds during summer \cite{hueg13,Reye15,Reye16,2017Moemken}.

These seasonal changes have strong implications not only on temperature and precipitation patterns, but also in the seasonal wind regimes and intra-annual variability. The seasonal variability of wind power generation increases under future climate conditions \cite{Reye15,hueg13,2017Moemken} even though the annual mean changes are comparatively small \cite{Tobi15,Tobi16,Reye16,hueg13,2017Moemken}. The impact may be large for the operational systems, and thus needs to be quantified adequately based on state-of-the-art climate model projections. For future research it would thus be highly desirable to provide larger ensembles of dynamically downscaled models for various regions on earth, possibly including wind speeds at hub heights.

\section*{Discussion} \label{sec:Discussion}

Wind power, PV and other renewable sources can satisfy the majority of the global energy demand \cite{Jacobson11,Sims11,Lu09}. However, system integration remains a huge challenge: The operation of wind turbines and PV relies on weather and climate and thus shows strong temporal fluctuations \cite{olau16,davi16,Heid10,Elsn15,diaz12,Schlachtb16,Bloo16,Hube14,Grams17}. The impact of climate change on the global energy yields of wind and solar power has been addressed previously \cite{Tobi15, Tobi16,Reye15,Reye16,2017Moemken,Jere15}, but the impact on fluctuations and system integration has remained unconsidered.

In this paper, we analyzed the change of the temporal characteristics of wind power generation in a strong (RCP8.5) and a medium climate change scenario (RCP4.5, see \cref{sec:a3}). Backup and storage needs increase in most of Central, Northern and North-Western Europe and decrease over the Iberian Peninsula, Greece and Croatia. As these effects are observed for both aggregation approaches used in this study (approach (a): aggregation per country, approach (b): European copperplate), we hypothesize that the effect will also be observed in intermediate scenarios with restricted interconnection between countries. Two main climatologic reasons for the observed increase were identified: a higher probability for long periods of low wind power generation and a stronger seasonal wind variability. The same tendencies, albeit at a different magnitude, are observed for different renewable penetrations, load time series and siting of wind turbines (see \cref{sec:a4}). 

The projected increase in backup energy needs may partly be compensated by using an appropriate mix of wind and PV. Furthermore, wind generation from offshore wind farms is often more persistent and installed capacities are strongly increasing. In a further study, climate projections for onshore and offshore wind and PV should thus be analyzed together in order to account for possible changes in the temporal variations of the combined system of renewables.

To isolate the change of the temporal characteristics of wind power generation, we made several simplifications. First of all, we assumed that wind provides a fixed share of the load for all time frames. This procedure normalizes out a possible change of global wind yields (already discussed in \cite{Tobi15, Tobi16,Reye15,Reye16,2017Moemken}). Technological progress of the wind turbines and changes of typical hub heights were not considered. For an integrated assessment, these aspects should be taken into account, but our approach reveals the impact of climate change on the temporal characteristics clearly.

A reliable interpretation of climate projections should be based on multi-model ensembles \cite{Tayl12,Tobi16}. Our analysis of the small EURO-CORDEX ensemble consisting of five models shows robust results regarding the sign of change for several regions in Europe. A statistical analysis of the output of 22 GCMs from the CMIP5 ensemble supports our findings, as the duration of periods with low values of the $f$-parameter over Central Europe is likely to increase. Large-scale climatologic developments leading to an increase of the seasonal wind variability were previously discussed in \cite{Demu09,Woll12,Zapp15,Lu07,Pint12,Fese15,hueg13,Reye15,2017Moemken}. For future research, it would be highly desirable if larger ensembles of dynamically downscaled models would be provided. Furthermore, data at turbine hub height should be made available. Ongoing downscaling experiments within the new CMIP6 CORDEX initiative \cite{Eyri16} will allow to assess the impact of climate change on system integration of intermittent renewables for various regions in the same manner. This should include a detailed and explicit analysis on the projected changes of both wind and PV. In conclusion, our work contributes to highlight the importance of integrated energy and climate research to enable a sustainable energy transition.

%
%
%
%
%

\section*{Acknowledgments}

We acknowledge the World Climate Research Programme's Working Group on
Regional Climate, and the Working Group on Coupled Modelling, former coordinating
body of CORDEX and responsible panel for CMIP5. We also thank the climate
modelling groups (listed in \cref{tbl:GCMs} in \cref{sec:a1}) for producing and making available
their model output. We also acknowledge the Earth System Grid Federation
infrastructure an international effort led by the U.S. Department of Energy's Program for
Climate Model Diagnosis and Intercomparison, the European Network for Earth System
Modelling and other partners in the Global Organisation for Earth System Science
Portals (GO-ESSP).
We thank H.~Elbern, G.~B.~Andresen, S.~Kozarcanin, T.~Brown, and J.~Hoersch for stimulating discussions. We gratefully acknowledge support from the Helmholtz Association (via the joint initiative ``Energy System 2050 -- A Contribution of the Research Field Energy'' and the grant no.~VH-NG-1025) and the Federal Ministry for Education and Research (BMBF grant no. 03SF0472) to D.~W.. For M.~R., J.~M.~and J.~G.~P.~contributions were partially funded by the the Helmholtz-Zentrum Geesthacht/German Climate Service Center (HZG/GERICS). J.G.P. thanks the AXA Research Fund for support. 

%


\appendix


\clearpage
\renewcommand{\theHfigure}{S\arabic{figure}} 
\setcounter{figure}{0}
\renewcommand{\thefigure}{S\arabic{figure}}
\renewcommand{\thetable}{S\arabic{table}}

\onecolumngrid
{\centering \Large  \textbf{Appendix}\\
accompanying the manuscript\\
\textbf{Impact of climate change on backup energy and storage needs in wind-dominated power systems in Europe\\}}
{\center \normalsize{Juliane Weber, Jan Wohland, Mark Reyers, Julia Moemken, Charlotte Hoppe, Joaquim G. Pinto and Dirk Witthaut}\flushright}
\quad\\
\quad\\


\section{Supporting material for the methods section} \label{sec:a1}

\begingroup
\setlength{\tabcolsep}{9pt} 
\renewcommand{\arraystretch}{1.5} 
\begin{table*}[htb]
\centering
\caption{{\bf Global circulation models used in the EURO-CORDEX ensemble.} 
The downscaled data has a resolution of 0.11$^\circ$ and 3 hours \cite{Jaco14}. Table adapted from \cite{Reye16}}
\label{tbl:GCMs}
\begin{tabular}{p{7.5cm}|p{7cm}}
\hline
\textbf{Model name}  & \textbf{Institution} \\
\hline
CNRM-CM5 (CNRM Coupled Global Climate Model, version 5) & Centre National de Recherches M\'{e}t\'{e}orologiques (CNRM), France  \\
\hline
EC-EARTH (EC-Earth Consortium) & European Consortium (EC) \\
\hline
IPSL-CM5A-MR (IPSL Coupled Model, version 5, coupled with NEMO, medium resolution) & Institut Pierre Simon Laplace (IPSL), France \\
\hline
MPI-ESM-LR (MPI Earth System Model, low resolution) & Max Planck Institute (MPI) for Meteorology, Germany \\
\hline
HadGEM2-ES (Hadley Centre Global Environment Model, version 2, Earth System) & Met Office Hadley Centre, United Kingdom \\
\hline
\end{tabular}
\end{table*}
\endgroup

\begin{figure}[h]
\centering
\includegraphics[width=0.5\textwidth]{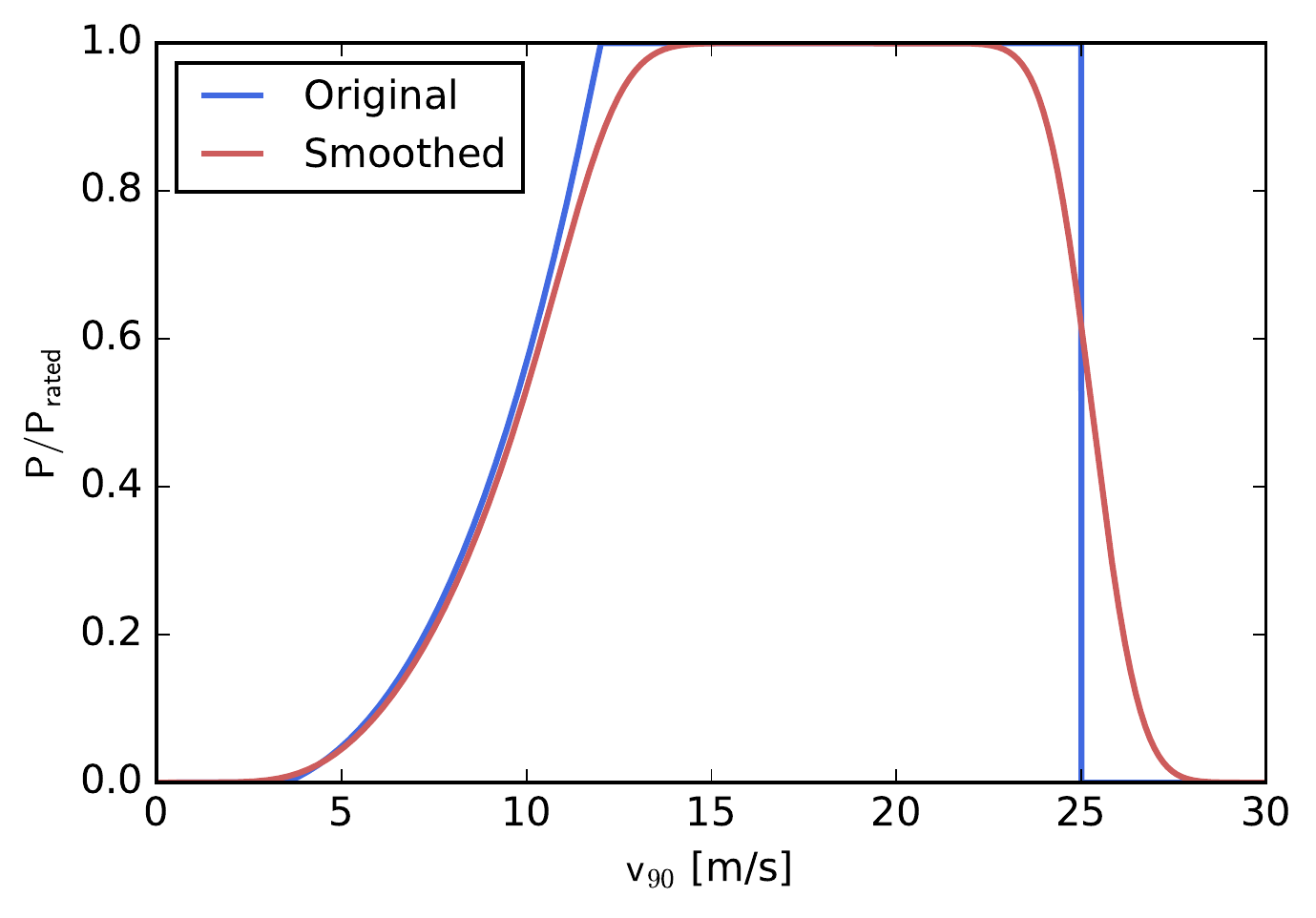}
\caption{
\label{fig:Power_Curve}
{\bf Power curve.} Power curve used to convert wind velocities in a height of 90 m ($v_{90}$) to wind power generation data. In order to account for wind farms and velocity variations the single turbine power curve (blue) is smoothed using a gaussian kernel (red).
}
\end{figure}

\begin{figure}[h]
\centering
\includegraphics[width=0.45\textwidth]{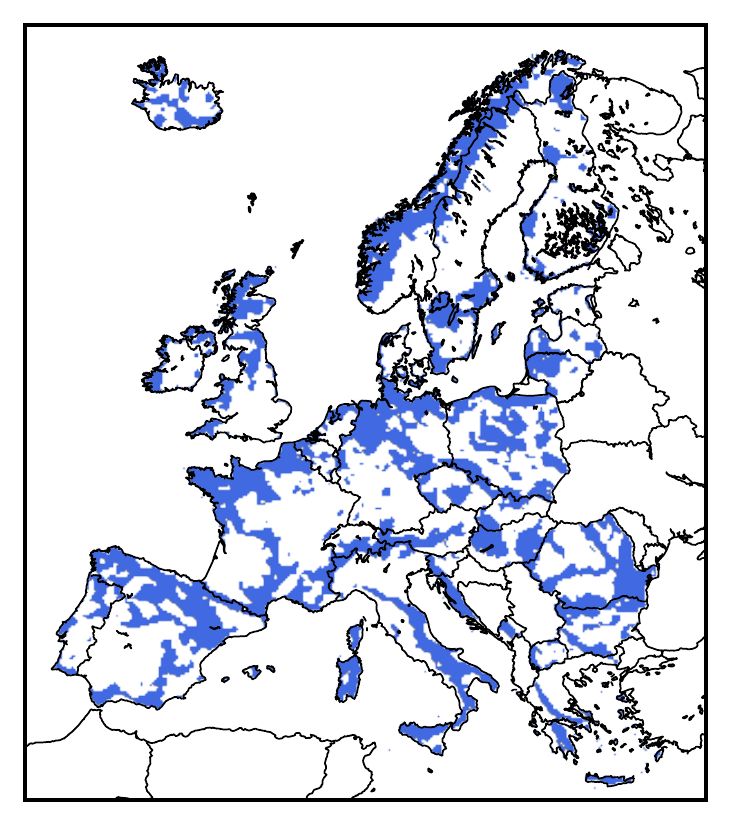}
\caption{
\label{fig:CF_weights}
{\bf Placement of wind farms.} 
Wind farms are homogeneously placed on colored sites. On these sites the 31-year average of the wind yield is higher than the country average. For this derivation, ERA-Interim data \cite{dee11} is used. In a sensitivity study, wind farms are placed homogeneously at each grid point inside a country (see \cref{fig:increase-backup_no_weights,fig:longcalms_no_weights,fig:seasonality_no_weights} in \cref{sec:a4}).
}
\end{figure}

\begin{figure}[h]
\centering
\includegraphics[width=0.45\textwidth]{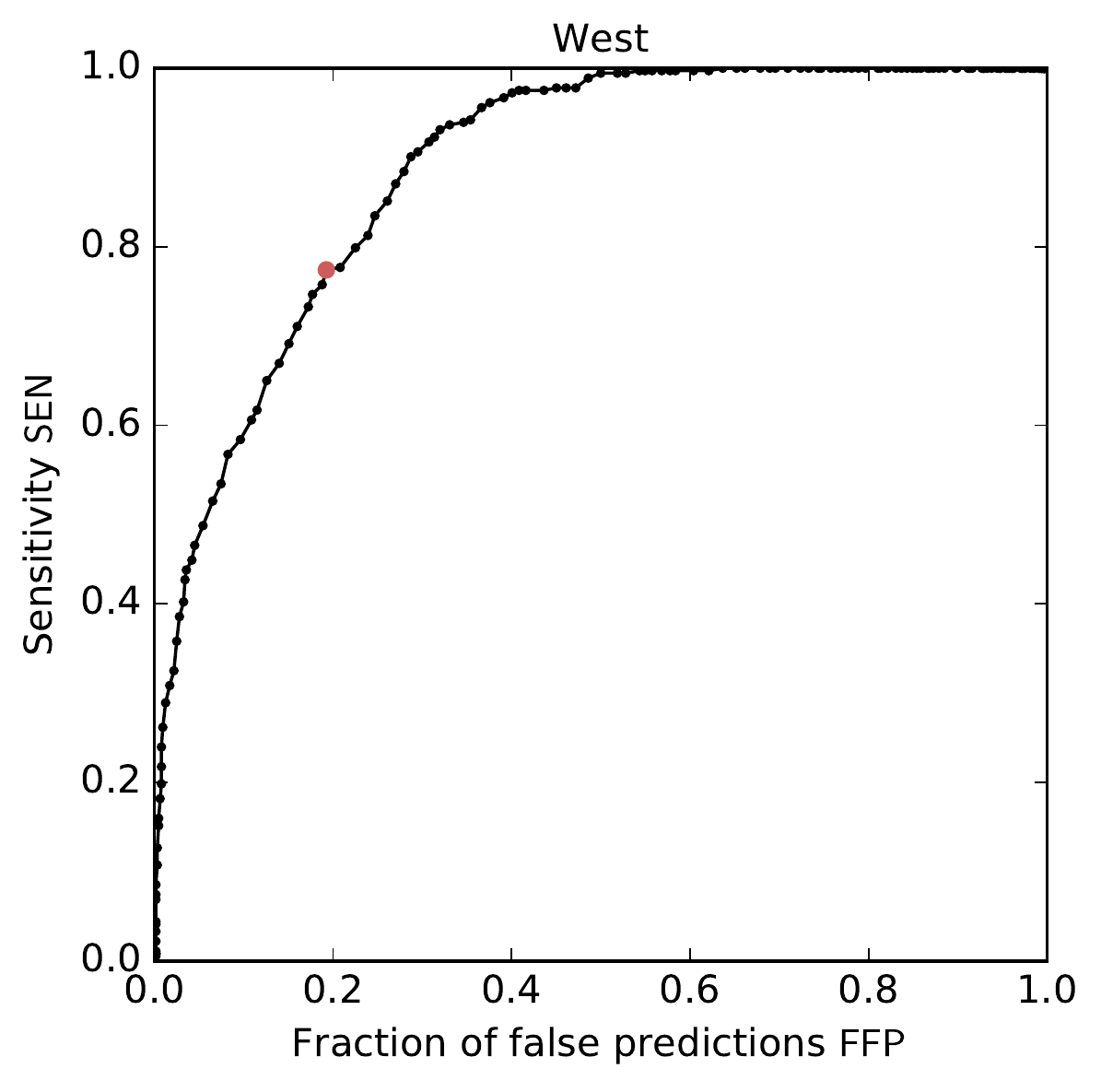}
\caption{
\label{fig:ROC}
{\bf ROC curve.} Receiver operating characteristic (ROC) curve to assess the performance of the $f$-parameter to classify days with low wind power generation.
For each pre-defined threshold value $f_{\rm th}$, the sensitivity ${\rm SEN}$ and the fraction of false predictions ${\rm FFP}$ are calculated. The optimal value of $f_{\rm th}$ (red dot) minimizes the distance to the point (${\rm SEN}$, ${\rm FFP}$) = (1,0), which corresponds to a perfect classifier. Results are shown for the western CWT.
}
\end{figure}

\clearpage

\section{Mid century (2030-2060)} \label{sec:a2}

In the main manuscript, most results are only shown for the end of the century (2070-2100). Here, results shall be shown for mid century (2030-2060).

\begin{figure*}[h]
\centering
\includegraphics[width=0.95\textwidth]{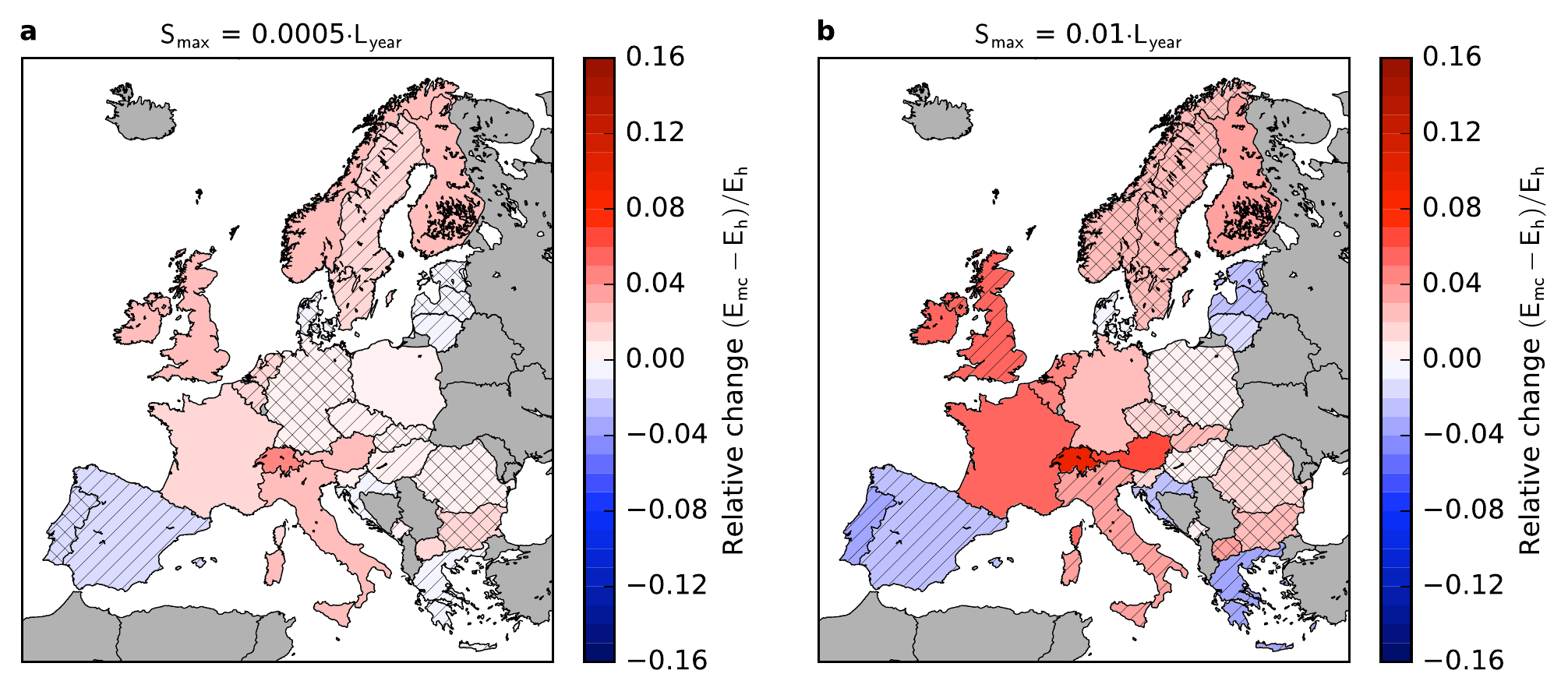}
\caption{
\label{fig:increase-backup_mc}
{\bf Impact of strong climate change on backup energy needs by mid century.} Parameters and presentation as in \cref{Fig2}.
}
\end{figure*}

In \cref{fig:increase-backup_mc} the relative change of the backup energy is shown for two different storage capacities. For most of Central Europe, France, the British Isles, Scandinavia (except Denmark) and Italy four to five models in the EURO-CORDEX ensemble predict an increasing backup need. Exceptions are Germany, where results are not robust for small $S_{\rm max}$ and Norway, Sweden, Poland and the Czech Republic, where results are not robust for high $S_{\rm max}$. For Spain, Greece and Croatia (and Portugal for high $S_{\rm max}$) four models predict a slightly decreasing backup need. Compared to the results we have for the end of the century (cf. Fig~2c and d in the main manuscript), changes are weaker and, therefore, in some countries less robust. For most countries in Eastern Europe results are not robust which is also observed for the end of the century.

\begin{figure*}[h]
\centering
\includegraphics[width=0.95\textwidth]{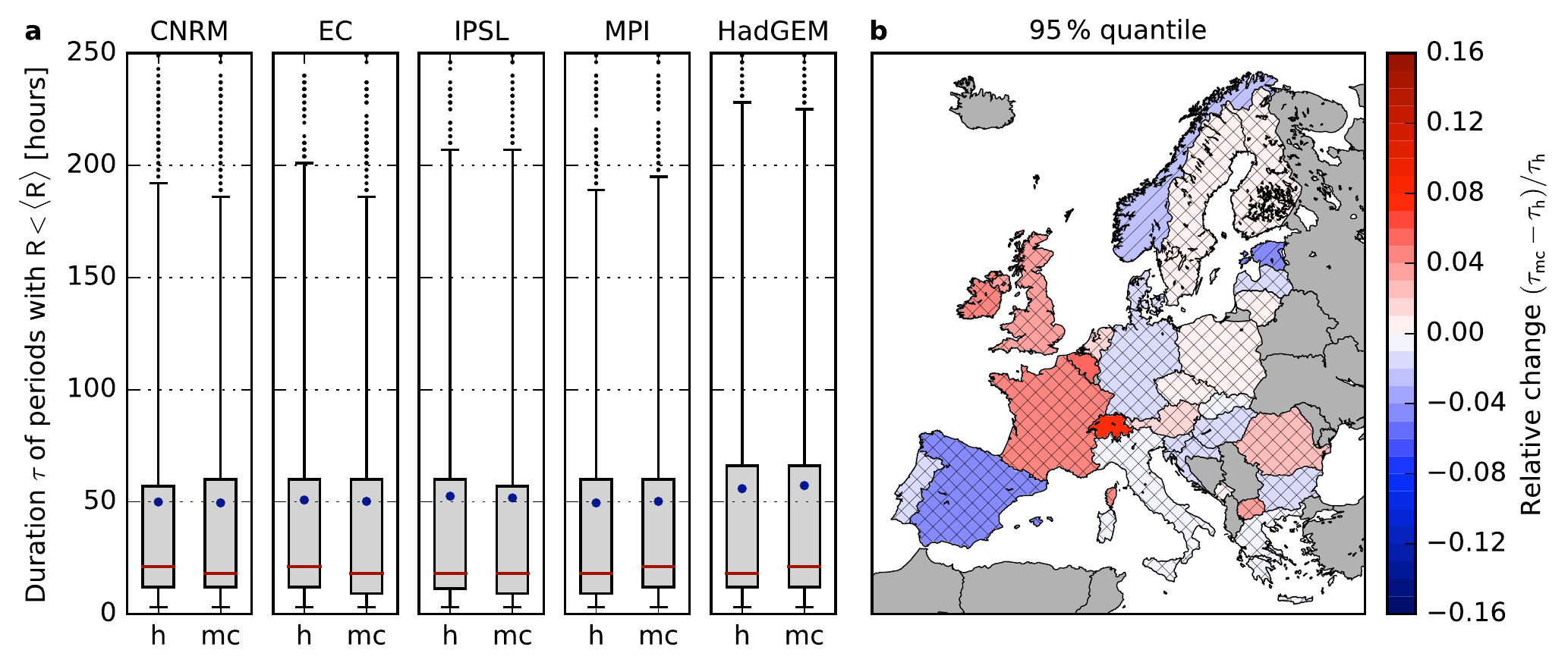}
\caption{
{\bf Change of the duration of periods with low wind generation by mid century.} 
Parameters and presentation as in \cref{Fig4}.
\label{fig:longcalms_mc}
}
\end{figure*}

The duration distribution of periods with $R(t) < \langle R \rangle$ and the relative change of the 95~\% quantile of this distribution are shown in \cref{fig:longcalms_mc}. In almost all European countries the five models do not agree on the sign of change. These results indicate that changes in the backup energy need can mostly not be attributed to a change in the duration of long low wind periods.

\begin{figure*}[h]
\centering
\includegraphics[width=0.45\textwidth]{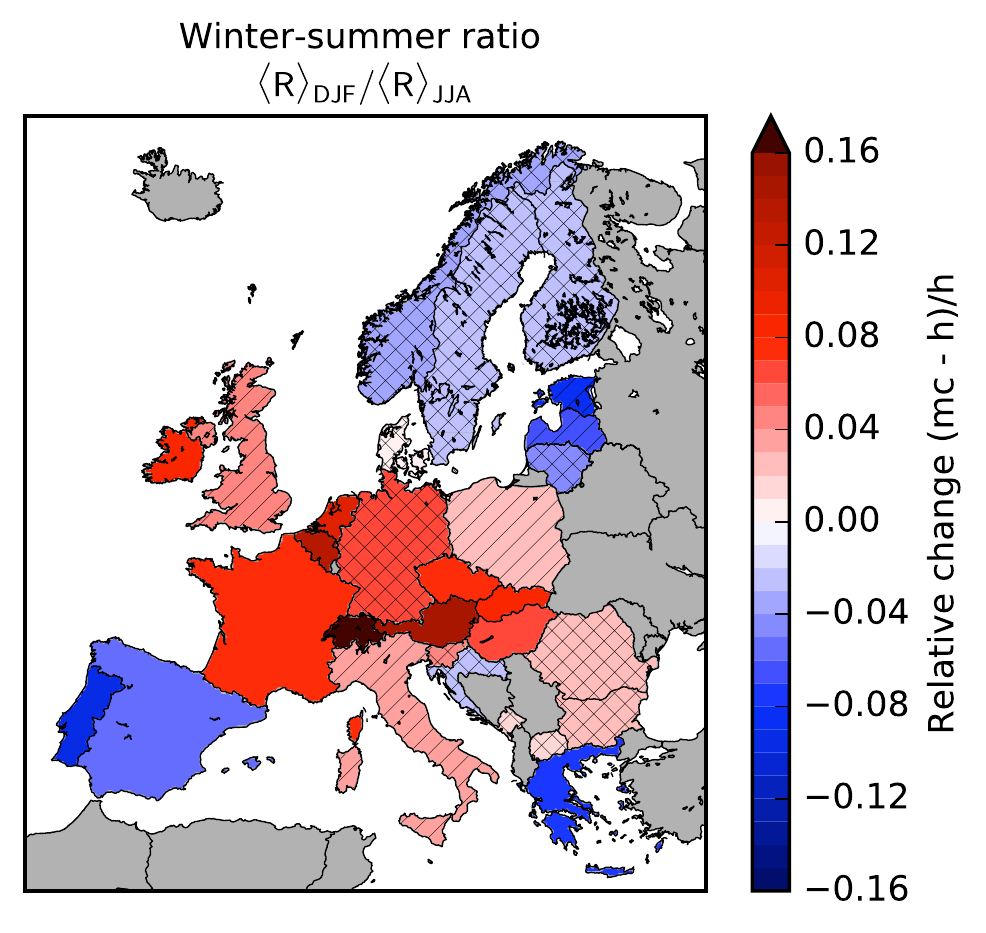}
\caption{
\label{fig:seasonality_mc}
{\bf Impact of strong climate change on the seasonal variability of wind power generation by mid century.} Parameters and presentation as in \cref{Fig6}.
}
\end{figure*}

Instead, higher/lower backup needs can be explained by the winter-summer ratio (see \cref{fig:seasonality_mc}) which is predicted to increase in France, the British Isles, and most of Central Europe (except Germany) by four to five models. Furthermore, the ratio is predicted to decrease on the Iberian Peninsula, Greece and (not robustly) in the Baltic countries. The relative change, however, is lower than by the end of the century (cf. Fig~6 in the main manuscript). 

Hence, by mid century, there is still a trend for increasing backup and storage needs due to a change in the seasonal wind variability. However, relative changes are weaker and results are less robust for some countries.

\clearpage

\clearpage

\section{RCP4.5} \label{sec:a3}

\begin{figure*}[h]
\centering
\includegraphics[width=0.95\textwidth]{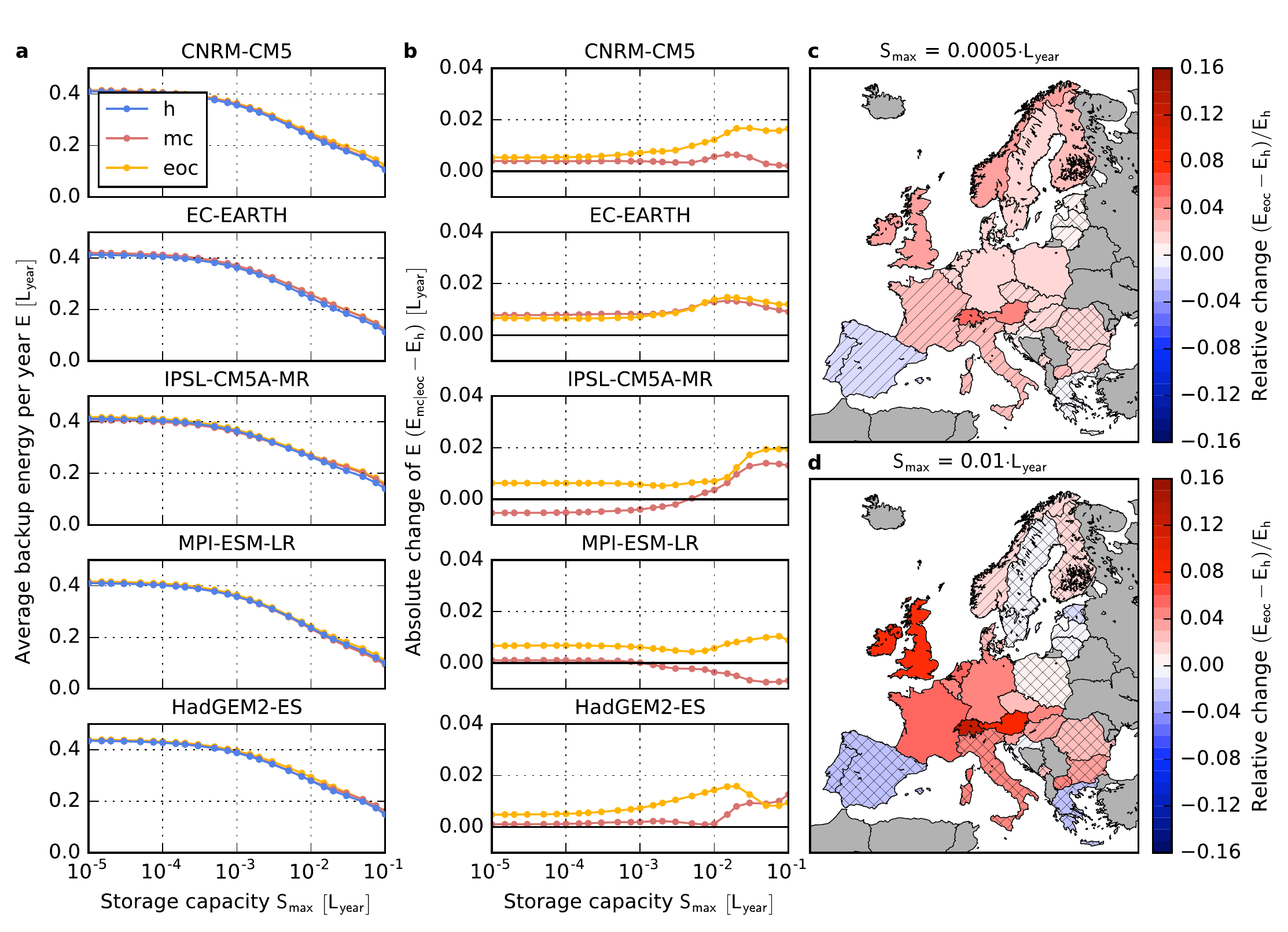}
\caption{
\label{fig:increase-backup_rcp45_eoc}
{\bf Impact of medium climate change on backup energy needs in Europe.} Parameters and presentation as in \cref{Fig2}.
}
\end{figure*}

We repeated the analysis for the medium climate change scenario RCP4.5 which is leading to an increase of 4.5~W/m$^2$ in radiative forcing ($\sim$650 ppm CO$_2$ equivalent) by 2100 \cite{Vuur11}. \Cref{fig:increase-backup_rcp45_eoc} shows the impact of this scenario on backup energy needs in Germany (panels a and b) and Europe (panels c and d) by the end of the century (2070-2100). The backup need increases in most of Central Europe, France and the British Isles. Even though the increase is weaker than in the RCP8.5 scenario (cf. Fig~2 in the main manuscript), it is important to note that the effect on the backup need is still pronounced in most of these countries. For France, Belgium, Scandinavia, Italy and Poland, the robustness of the results depends on the storage size. As for RCP8.5, results are not robust in most of Eastern Europe. The decrease in the backup need on the Iberian Peninsula (for high $S_{\rm max}$), Greece and Croatia is not robust.

\begin{figure*}[h]
\centering
\includegraphics[width=0.95\textwidth]{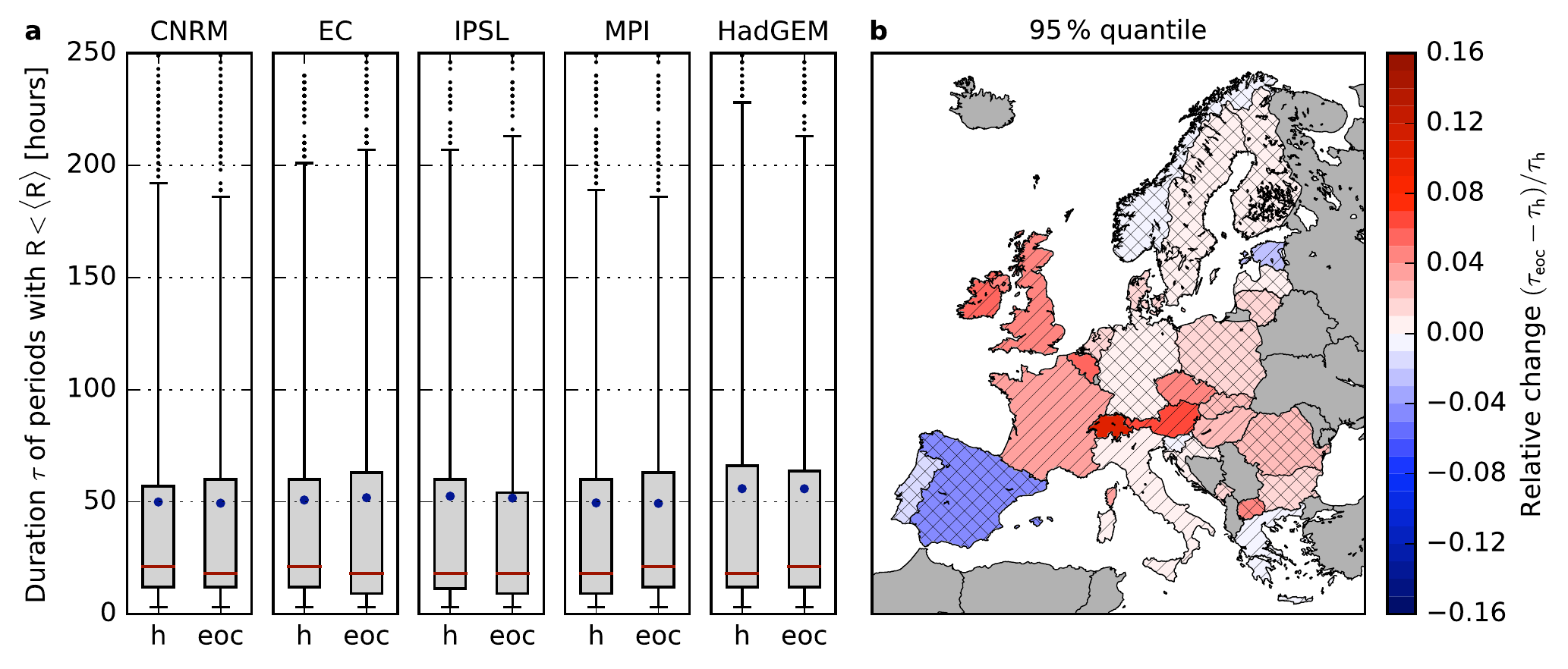}
\caption{
{\bf Change of the duration of periods with low wind generation for a medium climate change scenario.} 
Parameters and presentation as in \cref{Fig4}.
\label{fig:longcalms_rcp45_eoc}
}
\end{figure*}

The increase in the backup need can only partly be explained by an increase in the duration of long low wind periods (see \cref{fig:longcalms_rcp45_eoc}). Only in Switzerland and Belgium all models agree on the sign of change and on the British Isles, France, Austria and the Czech Republic, four of the five models agree on the sign of change. In the other countries, results are not robust and/or relative changes are small.

\begin{figure*}[h]
\centering
\includegraphics[width=0.45\textwidth]{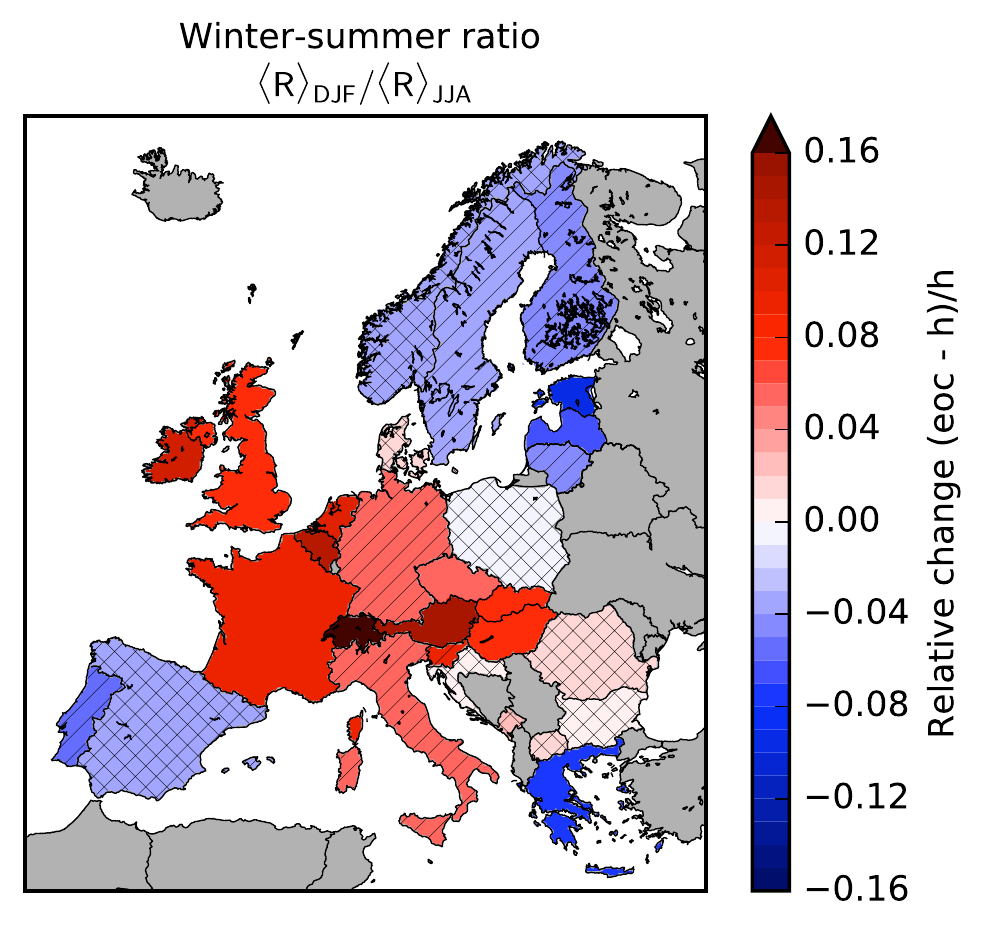}
\caption{
\label{fig:seasonality_rcp45_eoc}
{\bf Impact of medium climate change on the seasonal variability of wind power generation.} Parameters and presentation as in \cref{Fig6}.
}
\end{figure*}

Changes in the backup need can mostly be explained by changes in the winter-summer ratio (see \cref{fig:seasonality_rcp45_eoc}): For those countries, in which the backup energy need increases, the winter-summer ratio also increases (the British Isles, Benelux, France, Switzerland, Austria, Czech, Slovakia, Slovenia, Germany and Italy). 

In conclusion, by the end of the century similar trends as in the RCP8.5 scenario are found using the RCP4.5 scenario. However, changes are weaker and, therefore, often less robust. The decreasing backup needs on the Iberian Peninsula, Greece and Croatia are not robust for medium climate change.

\clearpage

\section{Sensitivity studies} \label{sec:a4}

In order to examine the sensitivity of our system on different parameters, we repeated our analysis using (a) different penetrations of renewables: $\gamma = 1.2$ and $\gamma = 0.8$, \cite{Rasm12}, (b) a constant load \cite{entsoe_load} and (c) a different distribution of wind farms. Furthermore, we repeated the analysis of the CMIP5 ensemble by using one threshold value of the $f$-parameter ($f_{\rm th}$) for all circulation weather types to characterize periods of low wind power generation.


\subsection*{Renewable penetration} \label{sec:gamma}

Different penetrations of wind power are simulated by scaling the renewable generation using the $\gamma$ factor: $\langle R \rangle = \gamma \, \langle L \rangle$. In the case $\gamma < 1$, on average an energy amount of $(1-\gamma)\cdot L_{\rm year}$ has to be provided by non-renewable power plants. Thus, we are interested in the additional backup energy $E_{\rm add} = E - (1-\gamma) \cdot L_{\rm year}$ with $E = \langle B \rangle / \langle L \rangle \cdot L_{\rm year}$.

\begin{figure*}[h]
\centering
\includegraphics[width=0.95\textwidth]{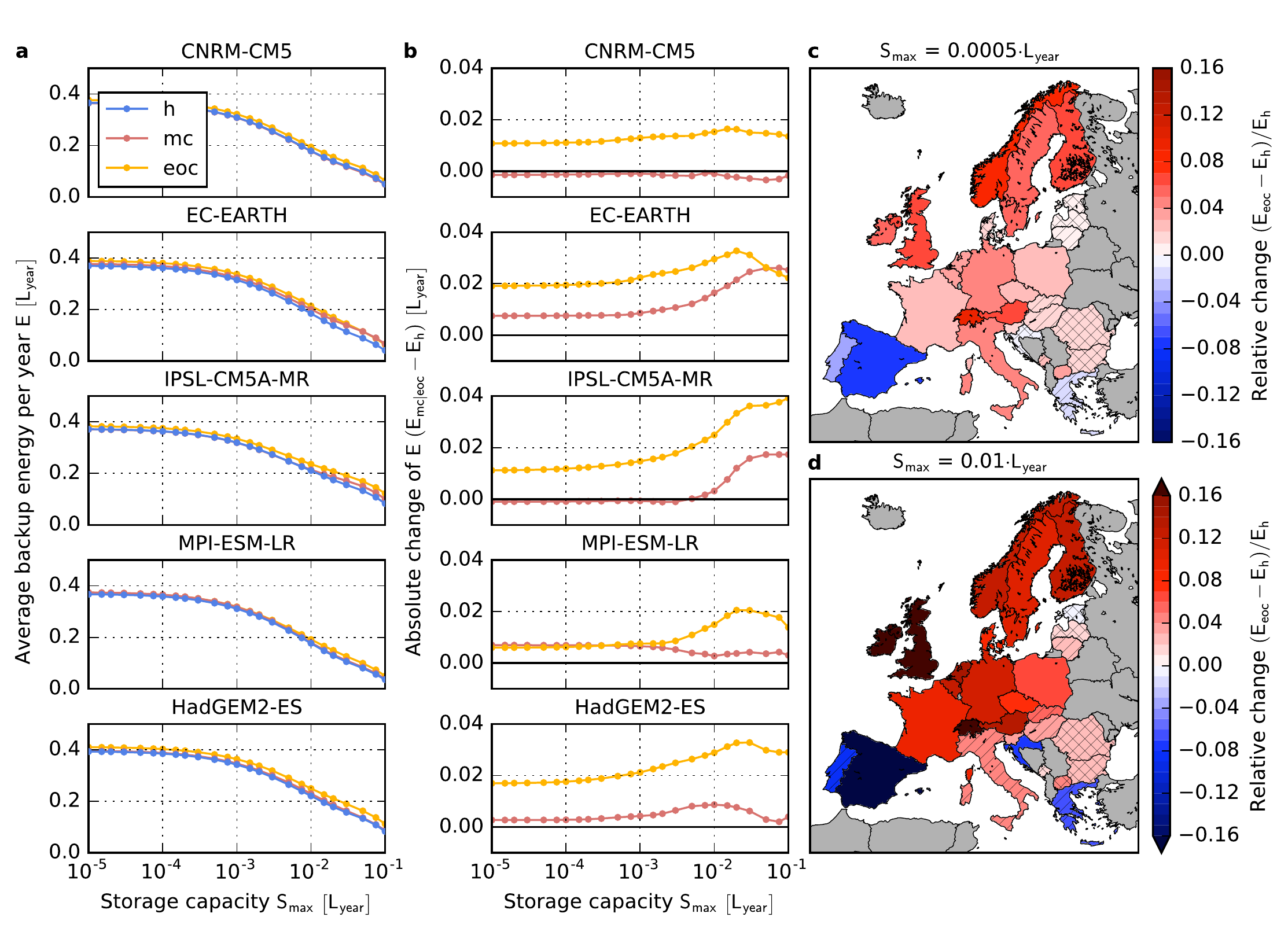}
\caption{
\label{fig:increase-backup_gamma120}
{\bf Impact of strong climate change on backup energy needs in Europe for a renewable penetration of $\mathbf{\boldsymbol{\gamma} = 1.2}$.} The maximum relative changes are 0.20 in Switzerland and -0.18 in Spain. Further parameters and presentation as in \cref{Fig2}.
}
\end{figure*}

\begin{figure*}[h]
\centering
\includegraphics[width=0.95\textwidth]{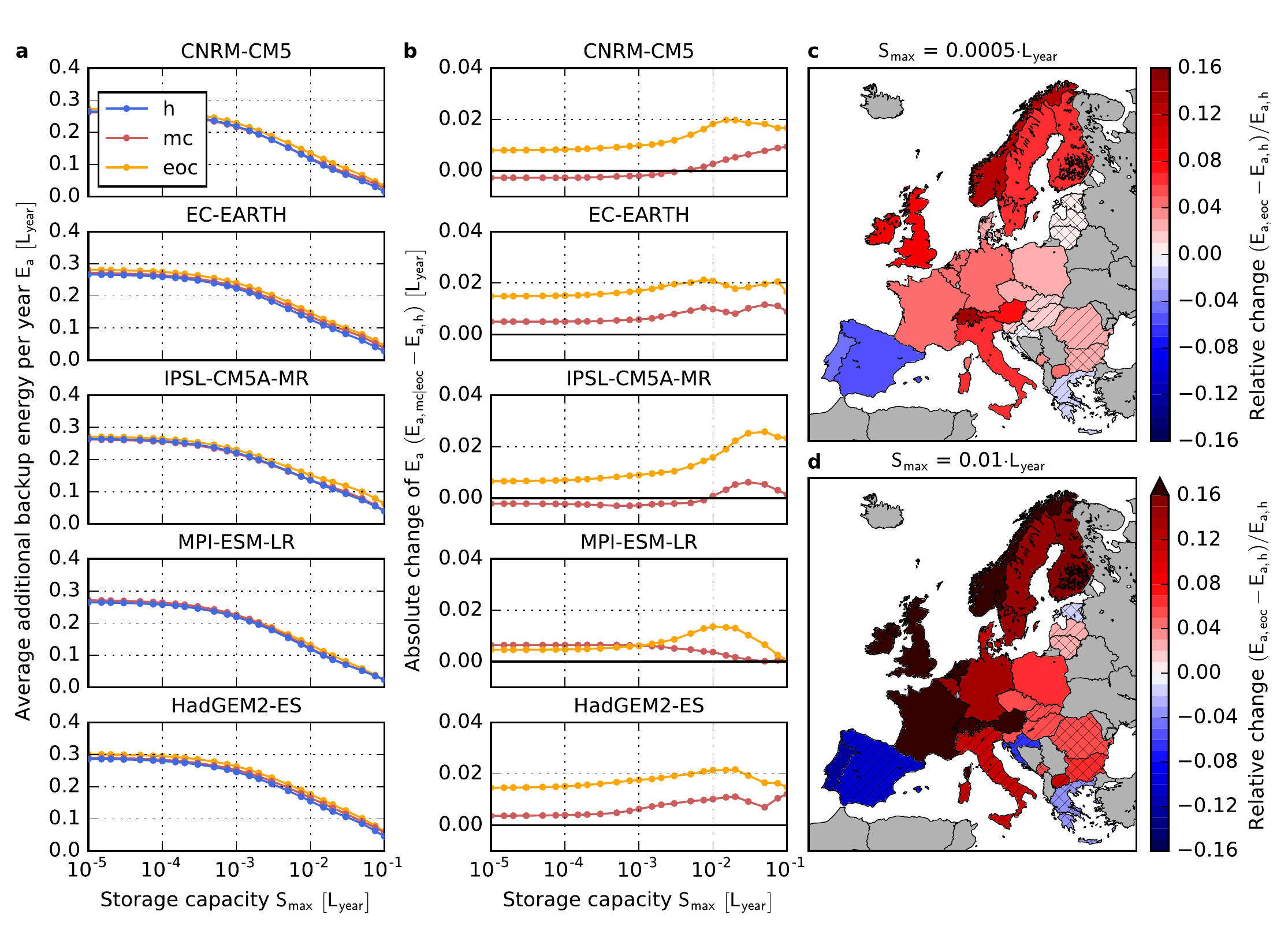}
\caption{
\label{fig:increase-backup_gamma80}
{\bf Impact of strong climate change on backup energy needs in Europe for a renewable penetration of $\mathbf{\boldsymbol{\gamma} = 0.8}$.} 
Shown is the amount of energy that has additionally to the pre-defined 20 \% to be provided by dispatchable backup generators. The maximum relative change is 0.36 in UK. Further parameters and presentation as in \cref{Fig2}.
}
\end{figure*}

In \cref{fig:increase-backup_gamma120,fig:increase-backup_gamma80} the change in the backup energy need is shown for $\gamma = 1.2$ and $\gamma = 0.8$, respectively. A comparison with Fig~2 of the main manuscript shows that absolute values of the backup energy (panels a) decrease (as expected). However, the absolute change of the backup energy (panels b) is quite similar and, thus, the relative change (panels c and d) is even higher for most countries (e.g. for UK the average relative change is 36.2~\% for $\gamma = 0.8$ and 20.2~\% for $\gamma = 1.2$ for $S_{\rm max} = 0.01 \cdot L_{\rm year}$). For all countries, the average predicted sign of change is the same as for $\gamma = 1$. Less models agree an the sign of change only in the case of Spain, Greece and Croatia (only high $S_{\rm max}$) for $\gamma = 0.8$ compared to $\gamma = 1$. 

Hence, a higher or lower renewable penetration also leads to increasing (decreasing) backup needs in many European countries, often even with higher relative changes.

\clearpage

\subsection*{Different load time series} \label{sec:load}

\begin{figure*}[h]
\centering
\includegraphics[width=0.95\textwidth]{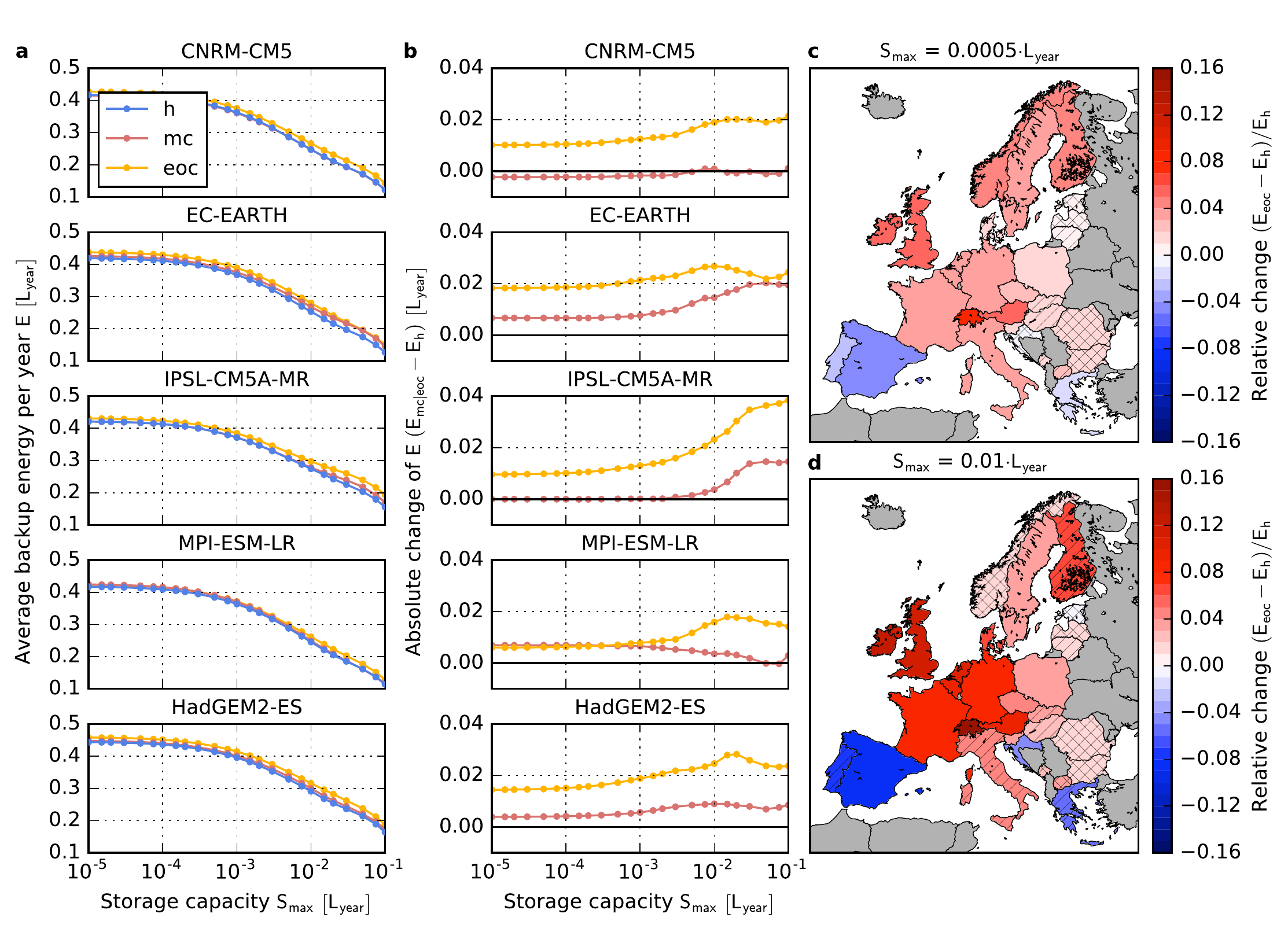}
\caption{
\label{fig:increase-backup_real_demand}
{\bf Impact of strong climate change on backup energy needs in Europe for a constant load.} Parameters and presentation as in \cref{Fig2}.
}
\end{figure*}

In order to analyze the sensitivity of the results on the applied load time series, we repeated our analysis using a constant value for the load such that: $L = \langle R \rangle$. The change of the backup energy need is shown in \cref{fig:increase-backup_real_demand}. A comparison with Fig~2 of the main manuscript reveals that the dependence of the results on the exact load time series is weak. Absolute values of the backup energy increase slightly (panels a). This comes from the fact that using a constant load leads to a loss of positive correlations: On average, there is more wind generation in winter than in summer. This correlates to a higher electricity demand in winter than in summer (in many countries). However, the relative change of the backup energy need is hardly affected. Only in Finland and Norway less models agree on the sign of change in the case of high $S_{\rm max}$.

Thus, we find only a weak dependence of the results on the exact load time series. This can further be explained by the fact that the fluctuations in the wind generation are much higher than fluctuations in the demand (cf. Fig~1c in the main manuscript) in scenarios with high wind penetrations.

\clearpage

\subsection*{Different distribution of wind farms} \label{sec:turbine_distribution}

In order to quantify the importance of the exact placement of wind farms, we repeated the analysis assuming a homogeneous distribution of wind farms at each grid point within a country.

\begin{figure*}[h]
\centering
\includegraphics[width=0.95\textwidth]{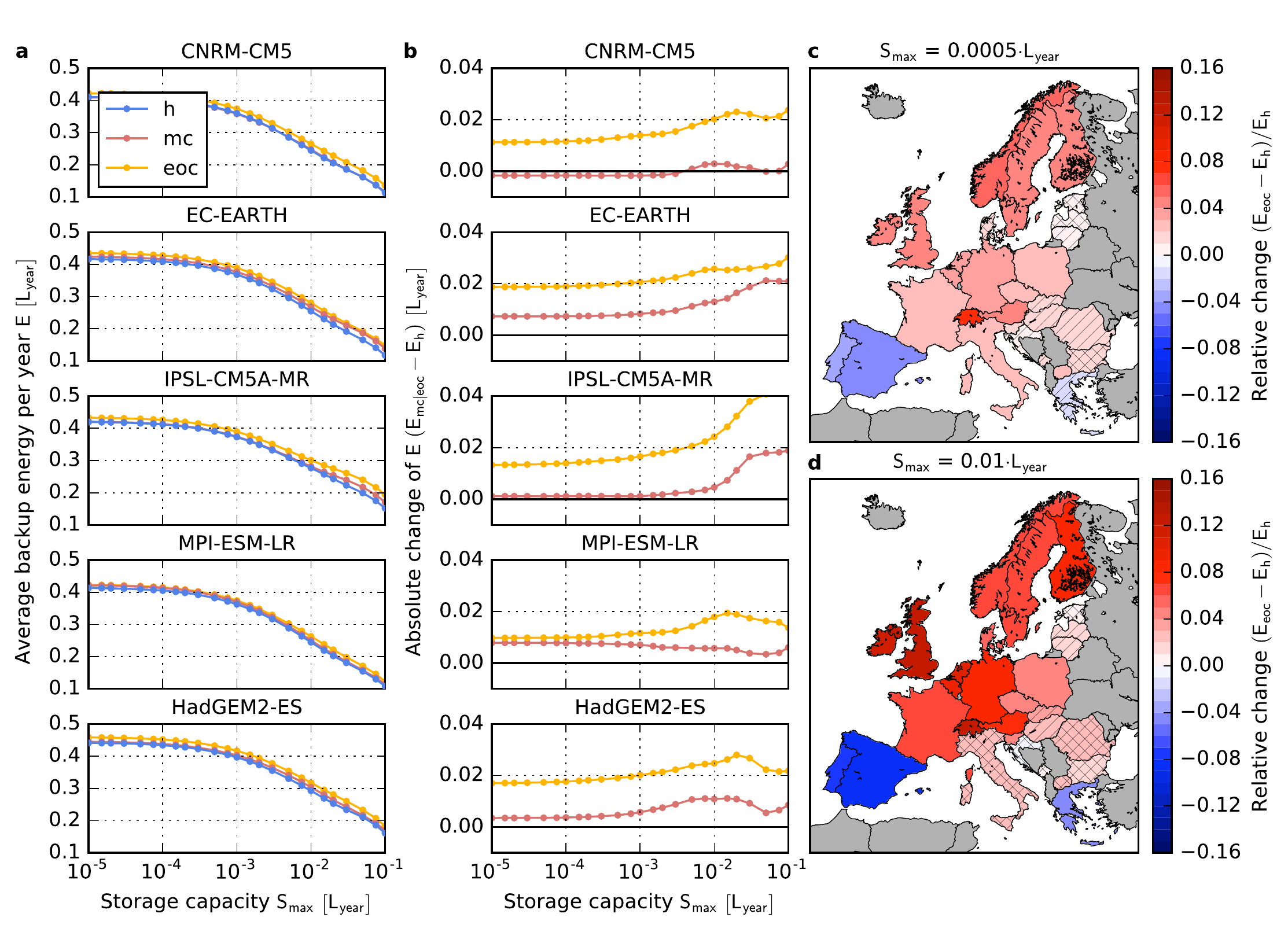}
\caption{
\label{fig:increase-backup_no_weights}
{\bf Impact of strong climate change on backup energy needs in Europe for a homogeneous distribution of wind farms.} Parameters and presentation as in \cref{Fig2}.
}
\end{figure*}

A comparison of \cref{fig:increase-backup_no_weights} with Fig~2 of the main manuscript shows that absolute backup energy values slightly increase (panels a). This is because wind farms are sited on less windy locations on average. However, the relative change of the backup energy hardly depends on the exact wind farm distribution. For Portugal and Greece results become more robust (for high $S_{\rm max}$) whereas in Croatia results become less robust.

\begin{figure*}[h]
\centering
\includegraphics[width=0.95\textwidth]{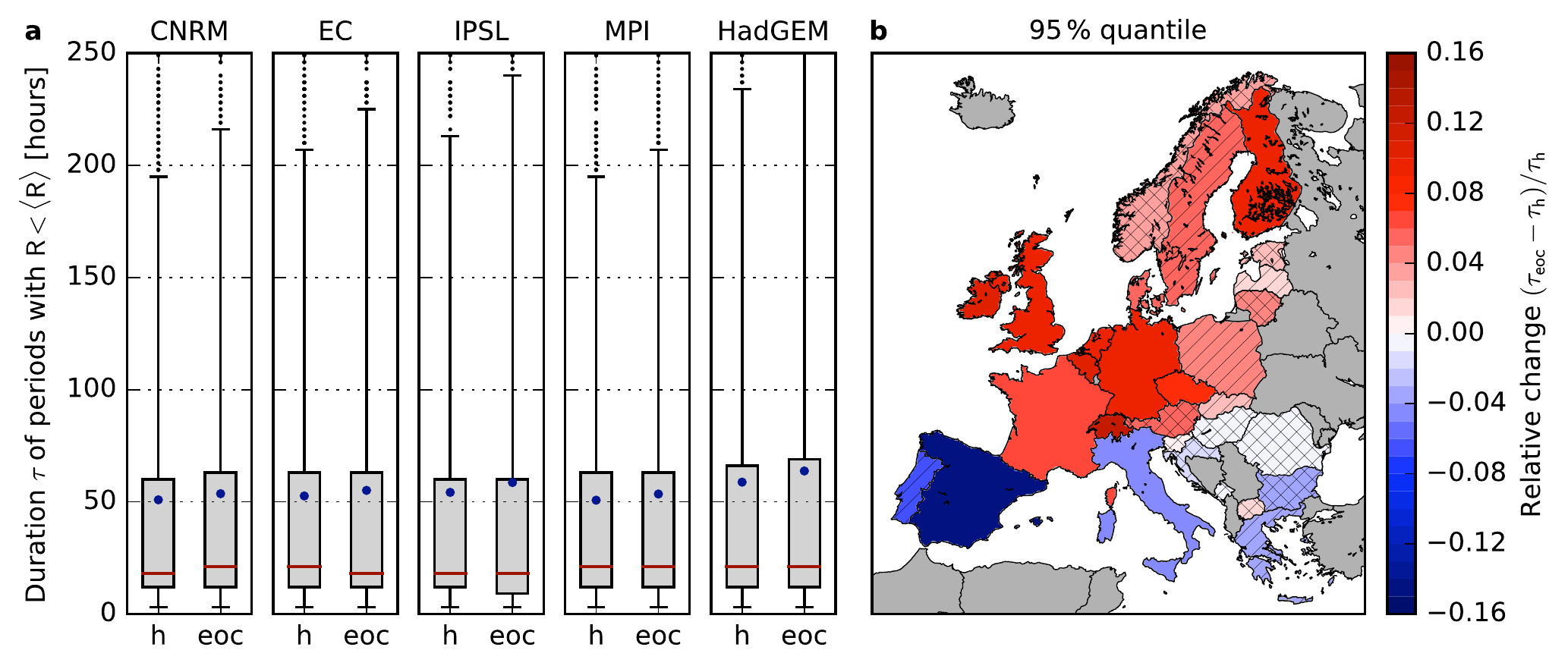}
\caption{
{\bf Change of the duration of periods with low wind generation for a homogeneous distribution of wind farms.} Parameters and presentation as in \cref{Fig4}.
\label{fig:longcalms_no_weights}
}
\end{figure*}

For Germany, the duration distribution of periods with $R(t) < \langle R \rangle$ is shifted to higher values for a homogeneous distribution of wind farms (cf. \cref{fig:longcalms_no_weights} and Fig~4 in the main manuscript, panels a) because wind farms are sited less favorable. Regarding the relative change of the 95 \% quantile of the duration distribution (panels b), results are less robust in Sweden, Austria, Estonia, Portugal and Slovenia whereas in Italy, Denmark, Greece and Latvia more models agree on the sign of change. In Italy, the sign of change flips -- the duration of long periods with low wind power output tends to decrease. In all other countries, especially in Central Europe, France and the British Isles the qualitative results are the same.

\begin{figure*}[h]
\centering
\includegraphics[width=0.45\textwidth]{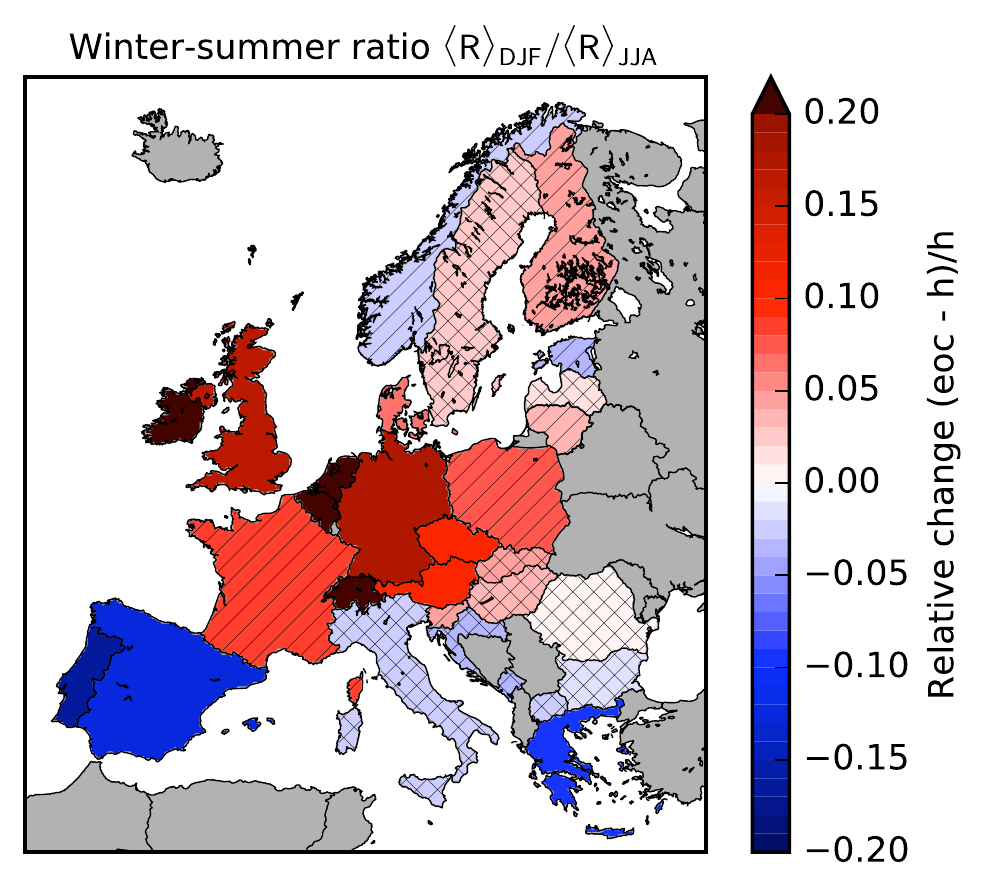}
\caption{
\label{fig:seasonality_no_weights}
{\bf Impact of strong climate change on the seasonal variability of wind power generation for a homogeneous distribution of wind farms.} Parameters and presentation as in Fig~6. Be aware that the scale of the colorbar is different than in \cref{Fig6}.
}
\end{figure*}

The relative change of the winter-summer ratio is more pronounced if wind farms are placed homogeneously on each grid point within a country (cf. \cref{fig:seasonality_no_weights} with Fig~6 in the main manuscript; be aware that the scales of the colorbars are different). The sign of change is the same in all countries. In Croatia and Slovakia results become less robust whereas in Finland and Lithuania more robust results are revealed.

All in all, the exact distribution of wind farms does hardly alter the results reported in this study (except for Italy).

\clearpage

\subsection*{Repetition of the analysis of the CMIP5 ensemble using a different threshold value of the $f$-parameter} \label{sec:f_thr_all}

\begin{figure*}[h]
\centering
\includegraphics[width=0.8\textwidth]{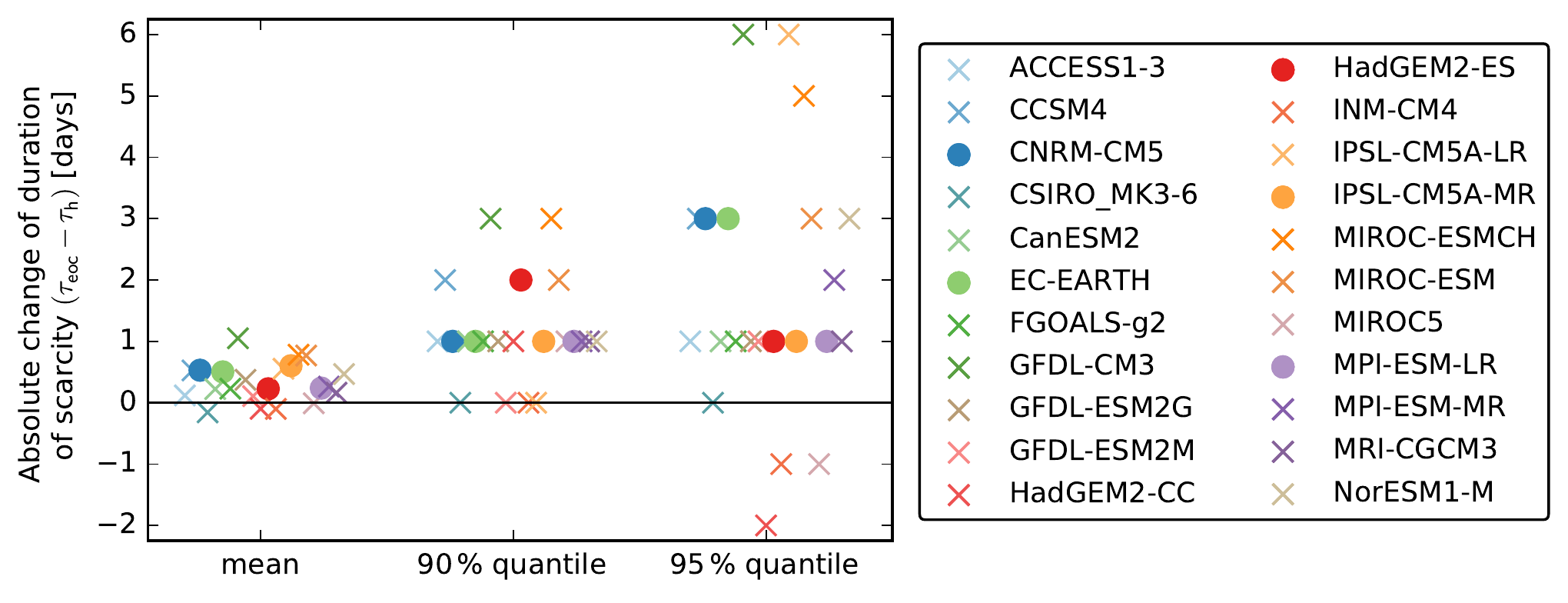}
\caption{
\label{fig:CWT-findings_one_f_thr}
{\bf Absolute change of the duration of periods with low wind generation using one $f_{\rm th}$ for all CWTs.} 
Parameters and presentation as in \cref{Fig7}b.
}
\end{figure*}

In the main manuscript, we determined the duration of periods with low wind generation by deriving one threshold value $f_{\rm th}$ for each CWT. It is important to note that the determined threshold value $f_{\rm th}$ is not perfect, as the distance $(1-{\rm SEN})^2 + {\rm FFP}^2$ is far away from being zero (see red dot in Fig C in S1\_Appendix). Hence, there is still a certain amount (for the western CWT: 19 \%) of false predictions. Thus, the derived duration distribution depends on the choice of the threshold value $f_{\rm th}$. To assess the sensitivity of the choice of $f_{\rm th}$, we repeat our analysis by determining one value for $f_{\rm th}$ which is independent of the underlying CWT. The result is shown in \cref{fig:CWT-findings_one_f_thr}. A comparison to Fig~7b in the main manuscript shows that in both cases the ensemble of the 22 GCMs predicts the same: The mean, the 90~\% quantile and the 95~\% quantile tend to increase by the end of the century. However, the value of a single model may be shifted for the 90~\% quantile and the 95~\% quantile: E.g. for HadGEM2-ES the absolute change of the duraton of scarcity of the 90~\% quantile is one day in Fig~7b and two days in \cref{fig:CWT-findings_one_f_thr}. An explanation for this is that already a slight shift of the threshold value $f_{\rm th}$ can split one long period with $f(t) \le f_{\rm th}$ into two shorter periods with $f(t) \le f_{\rm th}$ and a short period with $f(t) > f_{\rm th}$ in between or vice versa. Therefore, the results for the extremes (i.e. the 90~\% quantile and the 95~\% quantile) may be shifted for a single model. This effect averages out over the ensemble.

\end{document}